\documentclass[a4paper,11pt]{article}
\pdfoutput=1 

\usepackage{jcappub} 

\usepackage[T1]{fontenc} 

\title{\boldmath Constant-rate inflation: 
primordial black holes from conformal weight transitions
}

\author[a, b]{Kin-Wang Ng}
\author[c]{and Yi-Peng Wu}

\affiliation[a]{Institute of Physics, Academia Sinica, Taipei 11529, Taiwan} 
\affiliation[b]{Institute of Astronomy and Astrophysics, Academia Sinica, Taipei 11529, Taiwan} 
\affiliation[c]{Laboratoire de Physique Th\'{e}orique et Hautes Energies (LPTHE), \\
	UMR 7589 CNRS \& Sorbonne Universit\'{e}, 4 Place Jussieu, F-75252, Paris, France}

\emailAdd{nkw@phys.sinica.edu.tw}
\emailAdd{ywu@lpthe.jussieu.fr}

\abstract{
Constant-rate inflation, including ultra-slow-roll inflation as a special case, has been widely applied to the formation of primordial black holes with a significant deviation from the standard slow-roll conditions at both the growing and decaying phases of the power spectrum.
We derive analytic solutions for the curvature perturbations with respect to the late-time scaling dimensions (conformal weights) constrained by the dilatation symmetry of the de Sitter background and show that the continuity of conformal weights across different rolling phases is protected by the adiabatic condition of the inflaton perturbation. 
The temporal excitation of subleading states (with the next-to-lowest conformal weights), recorded as the ``steepest growth'' of the power spectrum, is triggered by the entropy production in the transition from the slow-roll to the constant-rate phases.
}

\definecolor{linkcolor}{RGB}{41, 127, 255}
\hypersetup{
	pdfstartview={XYZ},
	colorlinks,
	allcolors=linkcolor,
	bookmarksopen
}

\begin{document}
	\maketitle
	\flushbottom

\section{Introduction}\label{Sec. introduction}


Primordial black holes (PBHs) are interesting dark matter candidates that can be constrained by the ongoing and future detection of gravitational waves even if they only contribute a small fraction of the total dark matter density (see \cite{Carr:2020xqk,Green:2020jor} for recent reviews and references therein). That enhanced primordial density fluctuations from inflationary models with large curvature perturbations on small scales is the mainstream scenario for PBH formation. PBHs from models of inflation, with targeting masses spanning in the range of $10^{-17} - 10^9 M_\odot$, can be realized either in the single-field \cite{Cicoli:2018asa,Yokoyama:1998pt,Motohashi:2019rhu,Saito:2008em,Germani:2017bcs,Garcia-Bellido:2017mdw,Motohashi:2017kbs,Cheng:2018qof,Liu:2020oqe,Biagetti:2018pjj,Ballesteros:2020qam,Byrnes:2018txb,Bhaumik:2019tvl,Xu:2019bdp,Atal:2018neu,Kannike:2017bxn,Ragavendra:2020sop,Ozsoy:2018flq,Taoso:2021uvl} or multi-field \cite{Yokoyama:1995ex,Kawasaki:2012kn,Kawasaki:2016pql,Palma:2020ejf,Pi:2017gih,Clesse:2015wea,Fumagalli:2020adf,Cheng:2018yyr,Cheng:2016qzb,Anguelova:2020nzl,Braglia:2020eai,Fumagalli:2020nvq,Ozsoy:2020kat,Ozsoy:2020ccy} framework.  

In this work, we focus on the class of PBH models of inflation in the single-field framework with an approximately constant rate of rolling \cite{Motohashi:2014ppa,Martin:2012pe,Anguelova:2017djf,Tsamis:2003px,Kinney:2005vj}: $\delta \equiv \ddot{\phi} /(H \dot{\phi}) \approx \textrm{constant}$, where $\phi$ is the inflaton field and $H$ is the Hubble parameter during inflation. The analytic structure of the power spectrum generated by a constant-rate inflation has been investigated in \cite{Byrnes:2018txb,Liu:2020oqe,Cheng:2018qof,Ozsoy:2019lyy,Ballesteros:2020qam}, showing an essential departure from the standard slow-roll predictions \cite{Cicoli:2018asa} due to the significant violation of slow-roll conditions in the background dynamics \cite{Motohashi:2017kbs}. Quantum diffusion induced by the off-attractor nature of the constant-rate rolling can introduce non-negligible corrections on the large-scale spectrum via the stochastic effect \cite{Biagetti:2018pjj,Vennin:2020kng,Ando:2020fjm,Pattison:2021gpv,Pattison:2021oen,Ballesteros:2020sre}.

There are at least two model-independent features residing in the numerical results of the constant-rate models worthy of further clarifications, namely, better manifestations with very sharp transitions across different rolling phases and $\dot{\delta} \approx 0$ in each phase \cite{Cicoli:2018asa,Cheng:2018qof,Biagetti:2018pjj,Atal:2018neu,Liu:2020oqe}. First of all, the power spectrum $P_{\mathcal{R}}$ of the gauge invariant curvature perturbation $\mathcal{R}$ on superhorizon scales exhibits a $(k/k_\ast)^4$ scaling for $k \lesssim k_\ast$, where $k_\ast \approx -1/\eta_\ast$ is the horizon-crossing scale at which the inflation begins with a negative constant rate $\delta < -3/2$ (Phase 2 in Figure~\ref{fig:phase_plot}). Note that $\delta = -3$ is the so-called ultra-slow-roll limit \cite{Motohashi:2014ppa,Martin:2012pe,Tsamis:2003px,Kinney:2005vj}. Such a $(k/k_\ast)^4$ scaling of the power spectrum has been recognized in analytic approaches \cite{Byrnes:2018txb,Cheng:2018qof,Liu:2020oqe} and is an enhancement independent of the value of $\delta$. Here we refer such a $k^4$ enhancement as the ``steepest growth'' problem \cite{Byrnes:2018txb,Carrilho:2019oqg}. The $P_{\mathcal{R}}\sim k^4$ scaling sourced by sub-leading modes in $\mathcal{R}$ has been recognized in the ultra-slow-roll limit ($\delta = -3$) \cite{Carrilho:2019oqg}, and, in this work, we extend the argument to an arbitrary constant-rate inflation with $\delta \leq -3$.  
\footnote{A growth of the power spectrum faster than $k^4$ scaling (namely with a spectral index $n_s-1 >4$) can be realized in the multi-field inflationary models \cite{Braglia:2020taf,Fumagalli:2020adf,Fumagalli:2020nvq} or with an intermediate $\delta = -1/2$ phase between the primary slow-roll and the ultra-slow-roll phases in the single-field inflation \cite{Carrilho:2019oqg}.}

Second, the decay of the power spectrum from its maximal value exhibits a fixed scaling of the specific power, $P_{\mathcal{R}}\sim k^{6+2\delta}$, for a duration much longer than the period with a negative constant rate $\delta < -3$ (see \cite{Cicoli:2018asa,Cheng:2018qof,Biagetti:2018pjj,Atal:2018neu,Liu:2020oqe} for examples). Such a continuous scaling of the $k$-dependence in $P_{\mathcal{R}}$ would imply that the curvature perturbation $\mathcal{R}$ is not sensitive to the change of the inflaton potential shape in the post negative-constant-rate phase where $\delta$ may already become positive (Phase 3 in Figure~\ref{fig:phase_plot}). 
The fixed scaling $P_{\mathcal{R}}\sim k^{6+2\delta}$ derived by analytic approaches \cite{Byrnes:2018txb,Liu:2020oqe} in Phase 2 cannot be extrapolated to Phase 3, given that the ending time of Phase 2 is not a free parameter for arbitrary extension but is in fact constrained by the spectral amplitude of $P_{\mathcal{R}}$ for producing the desired PBH abundance. 
The fixed scaling of the power spectrum towards the end of inflation, regardless the change of $\delta$ from negative to positive values, is referred as the ``continuous scaling'' problem.   

In this work, we review both the steepest growth and continuous scaling problems from isometries of the de Sitter spacetime, having in mind that the scaling dimension of the late-time power spectrum (or the two-point correlator) is constrained by the dilatation symmetry between a rescaling of the spatial and temporal coordinates \cite{Antoniadis:2011ib,Creminelli:2011mw,Hinterbichler:2012nm,Maldacena:2011nz}. 
At a fixed time slice, these de Sitter invariant transformations project on the spatial hypersurface as conformal invariant operations, where the late-time correlation function $\langle \mathcal{R}(\vec{x}_1)\mathcal{R}(\vec{x}_2)\rangle \sim \vert \vec{x}_1-\vec{x}_2\vert^{-2\Delta}$ precisely describes a two-point conformal correlator with the scaling dimension (or conformal weight) $\Delta$ \cite{Antoniadis:2011ib,Arkani-Hamed:2018kmz,Arkani-Hamed:2015bza}.



The scaling dimensions/weights of a scalar field with a mass $m$ in $3+1$ dimensional de Sitter space is well known \cite{Strominger:2001pn,Ng:2012xp,Jafferis:2013qia}: It is a pair of real but non-integer numbers $\Delta_{\pm} = 3/2\pm \nu$ if $m/H < 3/2$, where $\nu = \sqrt{9/4 - m^2/H^2}$. In the late-time limit, only the branch led by the lowest power $\Delta_{-}$ survives. For a heavy field with $m/H > 3/2$, scaling dimensions become non-analytic (imaginary) and both $\Delta_{\pm}$ terms are important in the late-time correlators. Signatures of non-analytic momentum scaling due to the non-local exchange of heavy fields (including particles with non-zero spins) in curvature perturbations can be recorded in non-Gaussian correlators of $\mathcal{R}$ as oscillatory features \cite{Arkani-Hamed:2015bza,Arkani-Hamed:2018kmz,Chen:2009we,Chen:2009zp,Wu:2018lmx,Noumi:2012vr,Gong:2013sma,Lee:2016vti,Wang:2018tbf,Chen:2012ge,Pi:2012gf}.

On the other hand, the scaling dimensions of the late-time curvature perturbation in the constant-rate inflation is always analytic (real) but can be negative non-integer. As we will show that, during the $i$-th phase of inflation with the rate of rolling controlled by $\delta_i$, the leading power is $\Delta_i = 3/2 -\nu_i$ with (neglecting small corrections due to $\epsilon \equiv -\dot{H}/H^2$):
\begin{align}
\nu_i = \sqrt{\frac{9}{4} + 3\delta_i + \delta_i^2} = \left\vert \frac{3}{2} + \delta_i \right\vert,
\end{align}
which gives $\nu_i = 3/2+ \delta_i$ for $\delta_i > -3/2$ and $\nu_i = -\delta_i -3/2$ for $\delta_i < -3/2$. 

We revisit the steepest growth problem in Section~\ref{Sec. analytic_s} and the continuous scaling problem in Section~\ref{Sec. reduce_R} on the boundary surface at the transition time and show that both of them are consequences of the continuity of the scaling dimensions. Our arguments are supported by the bulk solutions in the de Sitter spacetime (Section~\ref{Sec. bulk12} and \ref{Sec. bulk3}). We show that the realization of PBH scenarios with enhanced power spectra led by a negative-constant-rate phase introduces yet unnoticed constraints on the scaling dimensions (see Section~\ref{Sec. PBH scenario}). In Section~\ref{Sec:criterion}, we review the entropy production of inflaton perturbations in PBH scenarios and clarify its connection with the conformal weight continuity. Finally, conclusions for the inflationary power spectrum characterizing with a time-varying conformal weight are given in Section~\ref{Sec. conclusion}.


\section{The (steepest) growth of power spectrum}\label{Sec. analytic_s}
\begin{figure}
	\begin{center}
		\includegraphics[width=12cm]{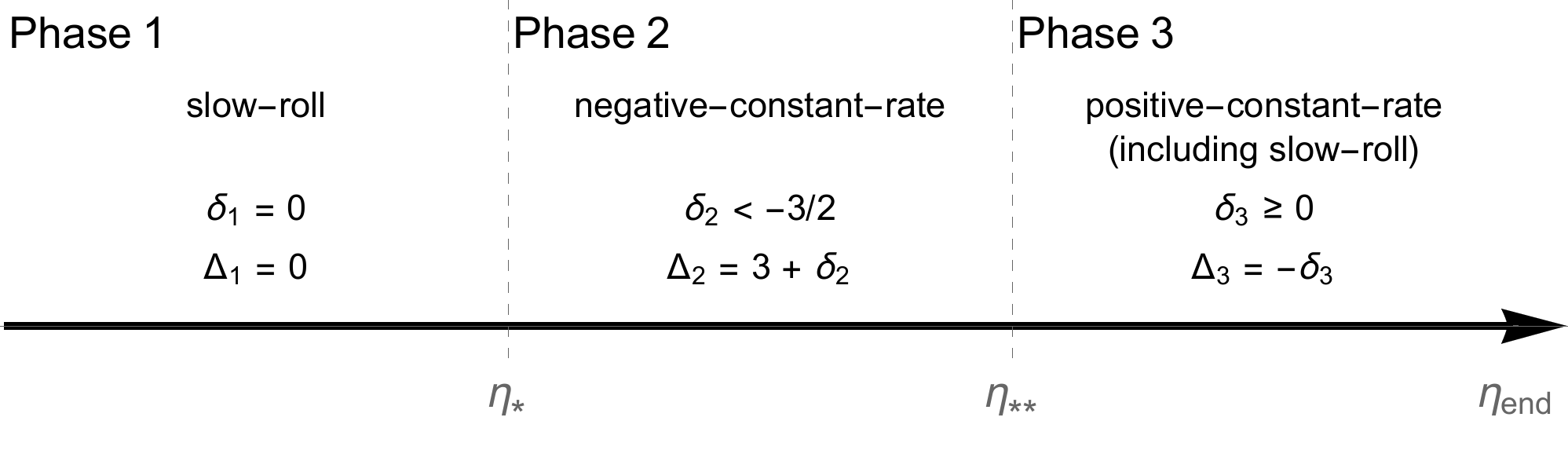}
	\end{center}
	\caption{A schematic description of the single-field inflation scenario considered in Section~\ref{Sec. analytic_s} and~\ref{Sec. reduce_R}, where $\delta_i$ is the $i$-th phase rate-of-rolling of inflaton and $\Delta_i$ is the scaling dimensionality (conformal weight) of the late-time curvature perturbation $\mathcal{R}$ in each phase. \label{fig:phase_plot}}
\end{figure}

In Section~\ref{Sec. analytic_s} and \ref{Sec. reduce_R} we consider a three-stage single-field inflation with sharp transitions to different phases at $\eta = \eta_\ast$ and $\eta = \eta_{\ast\ast}$, as given by Figure~\ref{fig:phase_plot}, where $\delta \equiv \ddot{\phi} /(H \dot{\phi})$ is the rate-of-rolling of inflaton $\phi$, and $\Delta \equiv 3/2 -\nu$ is the scaling dimension of the curvature perturbation $\mathcal{R}$ in the late-time limit 
\footnote{In this work, the late-time expansion of the mode functions $\mathcal{R}_k$ of the curvature perturbation is expanding with respect to the dimensionless time parameter $\tau \equiv -z \equiv -k\eta $ in the limit of $\tau \rightarrow 0$.}
 fixed by the dilatation symmetry of the four-dimensional de Sitter background \cite{Arkani-Hamed:2015bza,Arkani-Hamed:2018kmz,Antoniadis:2011ib}. Such a piecewise simplification of the inflationary dynamics can also be found in \cite{Ballesteros:2020qam,Ballesteros:2020sre,Taoso:2021uvl}. $\nu$ is the dimensionless mass parameter of the mode functions of $\mathcal{R}$. The first slow-roll parameters in Phase 2 and 3 are set to zero for simplicity (namely, $\epsilon_2 = \epsilon_3 = 0$).

\subsection{From slow-roll to negative-constant-rate}
Let us start with the background evolution of the first two stages of inflation, which are described by two sets of slow-roll parameters $\epsilon_1, \delta_1, \cdots$ for the physical time $t_0 \leq t < t_\ast$ and $\epsilon_2, \delta_2, \cdots$ for $t_\ast \leq t < t_{\ast\ast}$, where $t_\ast$ is the critical time for such a phase transition. We are interested in the cases where $\epsilon_1$ is constrained by CMB observations and $\epsilon_2 \ll \epsilon_1$ is very tiny and is responsible for the enhancement of the power spectrum associated with the PBH formation. Definition of the first slow-roll parameters are
\begin{align}
\epsilon = -\frac{\dot{H}(t)}{H^2(t)} = 1- \frac{\mathcal{H}^\prime(\eta)}{\mathcal{H}^2(\eta)} = 
\left\lbrace \begin{array}{cc}
\epsilon_1,   \quad \eta_0 &\leq \eta < \eta_\ast, \\
\epsilon_2,  \quad \eta_\ast &\leq \eta < \eta_{\ast\ast},
\end{array}\right.
\end{align}
where $\mathcal{H} \equiv a^\prime(\eta)/a(\eta) = a(t) H(t)$ is the conformal Hubble parameter in terms of the conformal time $d\eta = dt /a(t)$. 
The second slow-roll parameters are given by
\begin{align}
\epsilon_{i,2} \equiv \frac{\epsilon_i^\prime}{\mathcal{H}_i \epsilon_i} \approx 2\epsilon_i + 2\delta_i,
\end{align}
with $i = 1, 2$.
During the slow-roll phase (Phase 1), the derivative $\epsilon_1^\prime \sim \mathcal{O}(\epsilon_1^2) $ so that $\epsilon_1$ is treated as a constant for the background evolution. For $\eta > \eta_\ast$, inflation enters the phase with a constant rate of rolling $\delta_2 < -3/2$ and $\vert\delta_2 \vert \gg \epsilon_2$ significantly violates the slow-roll conditions. In general, $\epsilon_2$ is a function of $\eta$, yet for models of PBH formation $\epsilon_2$ becomes extremely tiny so that one can ignore its correction to the Hubble parameter $\mathcal{H}_2$.
The solutions of $\mathcal{H}$ in each stage are thus given by
\begin{align}
\mathcal{H}_1(\eta ) &= \frac{1}{1/\mathcal{H}_0 + (\epsilon_1 -1) (\eta - \eta_0)}, &&\quad \eta_0 \leq \eta < \eta_\ast,\\
\mathcal{H}_2(\eta ) &=\frac{1}{1/\mathcal{H}_\ast   -  (\eta - \eta_\ast) }, &&\quad \eta_\ast \leq \eta < \eta_{\ast\ast},
\end{align} 
where $\mathcal{H}_0\equiv \mathcal{H}(\eta_0 )$ and $\mathcal{H}_\ast \equiv \mathcal{H}(\eta_\ast ) = [ 1/\mathcal{H}_0 + (\epsilon_1 - 1)(\eta_\ast -\eta_0 ) ]^{-1}$.
Initial values $\mathcal{H}_0$ and $\eta_0$ at the beginning of inflation shall be regularized in such a way that is convenient for computing the physical observibles.
\footnote{
The conventional regularization of the Hubble parameter is to define $1/\mathcal{H}_0 \equiv -(1-\epsilon_1)\eta_0$ so that
\begin{align}
\label{Hubble_UV1}
\mathcal{H}_1(\eta ) &=- \frac{1}{(1- \epsilon_1 ) \eta}, &&\quad \eta_0 \leq \eta < \eta_\ast,\\ \label{Hubble_UV2}
\mathcal{H}_2(\eta ) &=\frac{1}{\epsilon_1 \eta_\ast -  \eta  }, &&\quad \eta_\ast \leq \eta < \eta_{\ast\ast}.
\end{align}
One can see that $\mathcal{H}_1$ takes the familiar form for the computation of the spectral index or the tensor-to-scalar ratio of the power spectrum in the framework of single-field inflationary models. However, such a regulation with respect to the slow-roll phase has a non-trivial limitation with the presence of a second-stage inflation, as one can see that the Hubble parameter \eqref{Hubble_UV2} may change into a negative value for $\eta >  \epsilon_1\eta_\ast$ before reaching the asymptotic boundary surface at $\eta = 0$.
}
Here we use a regularization with respect to the second-stage expansion as $1/\mathcal{H}_\ast = -\eta_\ast$. This gives the reduced expression as
 \begin{align}
 \label{Hubble_IR1}
 \mathcal{H}_1(\eta ) &= \frac{1}{-\epsilon_1\eta_\ast - (1- \epsilon_1 ) \eta}, &&\quad \eta_0 \leq \eta < \eta_\ast,\\ \label{Hubble_IR2}
 \mathcal{H}_2 (\eta )&=- \frac{1}{ \eta  }, &&\quad \eta_\ast \leq \eta < \eta_{\ast\ast}.
 \end{align} 
One can check that $\mathcal{H}$ is positively defined at all time. 

As a warm-up exercise, let us solve the mode functions of a massless scalar field $\phi$, which can apply to an inflaton in (ultra-)slow-roll with $\delta = 0$ (or $\delta = -3$). (The inflaton mode functions for general constant-rate cases is given in Section~\ref{Sec:criterion}). 
The equation of motion for a $k$-mode perturbation $\delta\phi_k$ in the Fourier expansion during inflation is
\begin{align}\label{eom:delta_phi}
\delta\phi_k^{\prime\prime} + 2 \mathcal{H}\delta\phi_k^\prime +k^2\delta\phi_k = 0,
\end{align}
where $\mathcal{H}$ is given by \eqref{Hubble_IR1} and \eqref{Hubble_IR2}. To obtain systematical expressions of the mode functions for different phases, it is convenient to use the rescaled variable $u_k \equiv a \delta\phi_k$ with respect to the dimensionless parameter $z \equiv k\eta$. The equation of motion for $u_k$ derived from \eqref{eom:delta_phi} reads 
\begin{align}
\frac{\partial^2 u_k}{\partial z^2} + \left(1 +  \frac{-\mathcal{H}^2 - \mathcal{H}^{\prime}}{k^2}\right) u_k = 0.
\end{align}
 We can further rewrite the Hubble parameter in the first and the second stages as $\mathcal{H}_1 = k/\tau_1$ and $\mathcal{H}_2 = k/\tau_2$, respectively, where we have defined $\tau_1 = -\epsilon_1z_\ast - r_1 z$ with $r_1 = 1-\epsilon_1$ and $\tau_2 = -r_2 z$ with $r_2 = 1-\epsilon_2$. As a result, the equations for $u_k$ become
\begin{align}\label{eom:massless_scalar}
r_1^2\, \frac{\partial^2 u_1}{\partial \tau_1^2} + \left(1+ \frac{\epsilon_1 - 2}{\tau_1^2}\right) u_1 &= 0, 
&&\quad \eta_0 \leq \eta < \eta_\ast,\\ \label{eom:massless_scalar2}
r_2^2\, \frac{\partial^2 u_2}{\partial \tau_2^2} + \left(1+ \frac{\epsilon_2 - 2}{\tau_2^2}\right) u_2 &= 0, 
&&\quad \eta_\ast \leq \eta < \eta_{\ast\ast}.
\end{align}
Here we keep $\epsilon_2$ in the equations for a comparison.

The general solution of \eqref{eom:massless_scalar} or \eqref{eom:massless_scalar2} takes the form of $\sqrt{\tau_i} H^{(1,2)}_{\nu_i}(\tau_i/r_i)$, where $H^{(1, 2)}_n(x)$ is the Hankel function of the first or second kind.
With respect to the standard Bunch-Davies vacuum in the UV limit ($-\eta\rightarrow \infty$), the coefficient for the $H_{\nu_1}^{(2)}$ term is set to zero, and the mode functions are given by
\begin{align}
u_1 &= c_1 \sqrt{\tau_1} H^{(1)}_{\nu_1} \left(\frac{\tau_1}{r_1}\right),  &&\quad \nu_1= \frac{1}{r_1}\sqrt{2-\epsilon_1 +\frac{r_1^2}{4}}, \\
u_2 &= c_2^{(1)} \sqrt{\tau_2} H^{(1)}_{\nu_2} \left(-z \right) + c_2^{(2)} \sqrt{\tau_2} H^{(2)}_{\nu_2} \left(-z \right) ,  &&\quad \nu_2= \frac{1}{r_2}\sqrt{2-\epsilon_2 +\frac{r_2^2}{4}}.  
\end{align}
In the limit of $\epsilon_i \rightarrow 0$, $\nu_i \rightarrow 3/2$ so that the scaling dimension of a massless field is known as $\Delta_i =3/2 -\nu_i \rightarrow 0$.
Note that $u_2$ is described by a generic (non Bunch-Davies) vacuum state due to the sharp transition from the slow-roll phase to the negative-constant-rate phase $\eta = \eta_\ast$. 
The coefficient $c_1$ can be fixed by the standard Bunch-Davies vacuum according to
\begin{align}
\lim_{-z \rightarrow \infty} u_1(\tau_1) \rightarrow c_1 \sqrt{-r_1 z} H^{(1)}_{\nu_1} (-z) = -\frac{i}{\sqrt{2k}} e^{-iz}.
\end{align} 
The large $x$ expansion of $H^{(1)}_{\nu_1}(x) = \sqrt{\frac{2}{\pi}}e^{-i\nu_1\pi/2-i\pi/4} e^{ix}/\sqrt{x} +\cdots $ gives
\begin{align}\label{c_1}
c_1 = -\frac{i}{2} \sqrt{\frac{\pi}{r_1 k}} e^{\frac{i}{2}(\nu_1+1/2)\pi}.
\end{align}
By matching $u(z)$ and $\partial u/\partial z$ at $z= z_\ast $, the coefficients $c_2^{(1)}$ and $c_2^{(2)}$ can be solved as
\footnote{A useful simplification, $( H^{(1)}_{\nu2 -1}(x) H^{(2)}_{\nu_2}(x) - H^{(1)}_{\nu_2} (x)H^{(2)}_{\nu_2 -1}(x) )^{-1} = -i\pi x/4$, has been used in the results of \eqref{c21_massless} and \eqref{c22_massless}.
}
\begin{align} 
\label{c21_massless}
c_2^{(1)} = \frac{i\pi}{4} c_1 &
 \left[-\tau_\ast H^{(1)}_{\nu_1-1}\left(\frac{\tau_\ast}{r_1}\right) H^{(2)}_{\nu_2}(\tau_\ast)  \right.\\\nonumber
&+ \left. H^{(1)}_{\nu_1}\left(\frac{\tau_\ast}{r_1}\right) \left(\tau_\ast H^{(2)}_{\nu_2-1}(\tau_\ast) +(A_2-A_1) H^{(2)}_{\nu_2}(\tau_\ast) \right) \right], \\
\label{c22_massless}
c_2^{(2)} = \frac{i\pi}{4} c_1  &
\left[ \tau_\ast H^{(1)}_{\nu_1 -1}\left(\frac{\tau_\ast}{r_1}\right) H_{\nu_2}^{(1)}(\tau_\ast)  \right. \\\nonumber
&+ \left. H^{(1)}_{\nu_1}\left(\frac{\tau_\ast}{r_1}\right) \left( -\tau_\ast H^{(1)}_{\nu_2-1}(\tau_\ast) +(A_1-A_2) H^{(1)}_{\nu_2}(\tau_\ast) \right) \right],
\end{align}
where $A_1 = r_1(1-2\nu_1)/2$ and $A_2 = r_2(1-2\nu_2)/2$.


We can compute the power spectrum of the massless scalar field according to the definition
\begin{align}
\left\langle \delta\phi_{\mathbf{k}} \delta\phi_{\mathbf{p}}\right\rangle 
= (2\pi)^3 \delta^{(3)}\left(\mathbf{k} + \mathbf{p}\right) P_{\delta\phi}(k; \eta)\frac{2\pi^2}{k^3},
\end{align}
where $\delta\phi_{\mathbf{k}} = \delta\phi_k a_{\mathbf{k}} +\delta\phi_k^\ast a_{-\mathbf{k}}^\dagger$ and $a_{\mathbf{k}}^\dagger $ ($a_{\mathbf{k}} $) is the creation (annihilation) operator of the free vacuum satisfying the commutation relation $[a_{\mathbf{k}}, a_{-\mathbf{p}}^\dagger] = (2\pi)^3 \delta^{(3)}(\mathbf{k} + \mathbf{p})$. 
The spectrum of $\delta\phi$ is led by
\begin{align}
P_{\delta\phi}(k; \eta) &= \frac{k^3}{2\pi^2} \frac{1}{a^2(\eta)} \left\vert u_k(\eta)\right\vert^2.
\end{align}
Here the scale factor $a(\eta)$ shall be expressed in terms of $a_k\equiv a(\eta_k)$ at which the $k$-mode exits the horizon. 
For $k < k_\ast = -1/\eta_\ast$ we can solve the scale factor $a(\eta)$ for the two stages via $\mathcal{H}$ given by \eqref{Hubble_IR1} and \eqref{Hubble_IR2} respectively, which indicates
\begin{align}
\frac{a_\ast}{a_k} &= \left[- \eta_\ast \mathcal{H}_k\right]^{-1/r_1}, \\
\frac{a}{a_\ast} &= \left(\frac{\eta}{\eta_\ast}\right)^{-1},
\end{align}
where $\mathcal{H}_k = \mathcal{H}(\eta_k) = a_k H_k = k$. The spectrum is therefore
\begin{align}
P_{\delta\phi}(k; \eta) = \frac{k^3}{2\pi^2} \left(\frac{a_\ast}{a(\eta)} \frac{1}{a_\ast}\right)^2 
\left\vert c_2^{(1)} \sqrt{-k\eta} H^{(1)}_{\nu_2}(-k \eta) +c_2^{(2)}\sqrt{-k\eta} H^{(2)}_{\nu_2}(-k \eta)\right\vert^2.
\end{align}
The small $x$ expansion of the Hankel functions are $H^{(1,2)}_n(x) = \mp\frac{i}{\pi} 2^n \Gamma(n) x^{-n} +\cdots $, and therefore 
for $k \ll -1/\eta$, the late-time limit of the power spectrum reads
\begin{align}
P_{\delta\phi}^{\rm IR}(k; \eta) = \frac{k H_k^2}{2\pi^2} \left(-z_\ast\right)^{\frac{2}{r_1}-2} \left(-z\right)^{3-2\nu_2} 
 \left\vert \left(c_2^{(2)}- c_2^{(1)}\right) \frac{i}{\pi} 2^{\nu_2}\Gamma(\nu_2) \right\vert^2,
\end{align} 
where both $c_2^{(1)}$ and $c_2^{(2)}$ are functions of $\tau_\ast = -z_\ast = k/k_\ast$, and
\begin{align}
c_2^{(2)}- c_2^{(1)} =& \frac{i \pi}{2} c_1 \tau_\ast 
	\left[ 
	J_{\nu_2}(\tau_\ast) H_{\nu_1-1}^{(1)}\left(\frac{\tau_\ast}{r_1}\right) \right.\\ \nonumber
	&\qquad \left. + H_{\nu_1}^{(1)} \left(\frac{\tau_\ast}{r_1}\right) 
	\left( \frac{A_1-A_2}{\tau_\ast}  J_{\nu_2} (\tau_\ast) -J_{\nu_2 -1} (\tau_\ast)  \right)
	\right].
\end{align}
In the case with $\epsilon_1 = \epsilon_2 =0$, $\nu_1 = \nu_2 =3/2$ and $c_2^{(2)}- c_2^{(1)} =  - c_1$. One can reproduce the standard result for a massless scalar as $P_{\delta\phi}^{\rm IR}(k; \eta) = H_k^2/(4\pi^2)$.



\subsection{Boundary arguments for steepest growth}\label{Sec. boundary12}
In this section we provide a simple argument to show that the enhancement of the curvature perturbation from slow-roll (Phase 1) to negative-constant-rate (Phase 2) is a $\mathcal{R} \sim k^2$ growth for a finite range $k_{\rm min} < k < k_\ast$ for modes exit the horizon close to the end of Phase 1 \cite{Byrnes:2018txb,Liu:2020oqe}.
Such a growth that corresponds to a power spectrum with the spectral index $n_s -1 =4$ is recognized as the steepest possible growth beyond the limitation of the ordinary matter power spectrum \cite{Byrnes:2018txb}. Our arguments only base on the continuity of matching the late-time curvature perturbations, $\mathcal{R}_1$ and $\mathcal{R}_2$, at the boundary surface $\eta = \eta_\ast$ with their leading dimensionality, $\Delta_1$ and $\Delta_2$, constrained by the dilatation symmetry of the de Sitter background \cite{Arkani-Hamed:2015bza,Arkani-Hamed:2018kmz}. 

The first step is to find out the correct scaling dimension of $\mathcal{R}$ in each phase. The dynamics of the curvature perturbation $\mathcal{R}$ can be computed by virtue of the Mukhanov-Sasaki variable $v = -y \mathcal{R}$, where $y = a \sqrt{\epsilon}$ includes the first slow-roll parameter and each Fourier mode function $v_k$ follows the equation
\begin{align}\label{eq:Mukhanov_sasaki}
v_k^{\prime\prime} + \left( k^2 - \frac{y^{\prime\prime}}{y}\right) v_k =0.
\end{align}
The ratio $y^{\prime\prime}/y = \mathcal{H}^2 (2+ 2 \epsilon + 3\delta + \delta^2 + \cdots)$ features the time-varying mass term for the mode function.
To identify the steepest growth, it is enough to impose $\epsilon_1 = \delta_1 = \epsilon_2 =0$ for the computation of $v_k$, which gives $y_1^{\prime\prime}/y_1 = 2\mathcal{H}_1^2$ and $y_2^{\prime\prime}/y_2 = \mathcal{H}_2^2(2+3\delta_2 +\delta_2^2)$. Note that such a simplification also makes $\mathcal{H}_1 = -1/\eta = \mathcal{H}_2$ and thus \eqref{eq:Mukhanov_sasaki} reads
\begin{align}\label{eom:MS_1}
\frac{\partial^2 v_1}{\partial \tau_1^2} + \left(1+ \frac{\nu_1^2 - 1/4}{\tau_1^2}\right) v_1 &= 0, 
&&\quad \eta_0 \leq \eta < \eta_\ast,\\ \label{eom:MS_2}
 \frac{\partial^2 v_2}{\partial \tau_2^2} + \left(1+ \frac{\nu_2^2 - 1/4}{\tau_2^2}\right) v_2 &= 0, 
&&\quad \eta_\ast \leq \eta < \eta_{\ast\ast},
\end{align}
where $\tau = -z = -k\eta$. One can observe that $\nu_1 = 3/2$ and $\nu_2 = (9/4 +3\delta_2 +\delta_2^2)^{1/2} = \vert 3/2 + \delta_2 \vert$, which implies that $\Delta_1 = 0$ and $\Delta_2 = 3/2 -\nu_2 = 3 +\delta_2$ for $\delta_2 < -3/2$. 

Given that $\Delta_1$ and $\Delta_2$ are only real numbers, we can expand the curvature perturbation $\mathcal{R}_k = -v_k/y$ in the limit of $\tau = -k\eta \rightarrow 0$ as
\begin{align}\label{def:late_time_R}
k^{3/2}\mathcal{R}_i \rightarrow \mathcal{C}_{i0} (-k\eta)^{\Delta_i} + \mathcal{C}_{i 1} (-k\eta)^{\Delta_i + 2} + \cdots,
\end{align}
where $\mathcal{C}_{i n}$ are late-time coefficients for the power series of $(-k\eta)^{\Delta_i + 2n}$.
\footnote{The even power $2n$ of this expansion comes from the power series expression of the Hankel function, where
\begin{align}\nonumber
H_\nu^{(1)}(x) = \sum_{n = 0}^{\infty} \frac{(-1)^{n}}{n!} \left(i \csc \pi \nu \right)\left(\frac{x}{2}\right)^{2n} 
\left[e^{-i\pi \nu}  \frac{1}{\Gamma(n+1 + \nu)} \left(\frac{x}{2}\right)^{\nu}-  \frac{1}{\Gamma(n+1 - \nu)}\left(\frac{x}{2}\right)^{-\nu}\right],
\end{align}
and $H_\nu^{(2)}(x)$ is the complex conjugate if both $\nu$ and $x$ are real.
}
 The coefficients $\mathcal{C}_{i n}$ in the definition of \eqref{def:late_time_R} are dimensionless but can have $k$-dependence, and the combination $\mathcal{C}_{i n} k^{\Delta_i +2n}$ can be taken as primary operators of a conformal field theory with dimension $\Delta_i +2n$. In particular, $\mathcal{C}_{1n}$ are fixed by $c_1$ in \eqref{c_1} from the Bunch-Davies vacuum, and the explicit form of $\mathcal{C}_{2n}$ can be found in the next section.  For the late-time expansion to be valid in Phase 1 the momentum $k$ shall satisfy $\tau_\ast\equiv -k\eta_\ast = k/k_\ast \ll 1$.

There are two boundary conditions come from the continuity of $\mathcal{R}$ at $\eta =\eta_\ast$. The first condition $\mathcal{R}_1(\eta_\ast) =\mathcal{R}_2(\eta_\ast)$ gives
\begin{align}\label{boundary_constraint_1}
\mathcal{C}_{10} \left(\frac{k}{k_\ast}\right)^{\Delta_1} + \mathcal{C}_{11} \left(\frac{k}{k_\ast}\right)^{\Delta_1 +2} +\cdots
= \mathcal{C}_{20} \left(\frac{k}{k_\ast}\right)^{\Delta_2}, 
\end{align}
which implies that the leading contribution to the coefficient $\mathcal{C}_{20}$ in the limit of $k/k_\ast \ll 1$ is given by the term with the lowest power and thus $\mathcal{C}_{20} \sim (k/k_\ast)^{\Delta_1 - \Delta_2}$. As a result, the late-time curvature perturbation obtained in Phase 2 reads $k^{3/2}\mathcal{R}_2\sim k^{\Delta_1}$, which is independent of $\Delta_2$. This asserts that the very large-scale modes (well outside the horizon during Phase 1) are not affected by a subsequent phase transition at late times. 

However, for the special case $\Delta_1 \rightarrow 0$, the subleading terms $\mathcal{C}_{1n}$ with higher powers ($n >0$) can become important when $k/k_\ast \rightarrow 1$ and change the momentum dependence of $\mathcal{C}_{20}$ .
This fact is manifest from the first derivative $\mathcal{R}_1^\prime(\eta_\ast) =\mathcal{R}_2^\prime(\eta_\ast)$ as
\begin{align}\label{boundary_constraint_2}
A_1 \mathcal{C}_{10} \left(\frac{k}{k_\ast}\right)^{\Delta_1-1} + \left(A_1 \mathcal{C}_{11} +B_1\right) \left(\frac{k}{k_\ast}\right)^{\Delta_1 +1} +\cdots
=A_2  \mathcal{C}_{20} \left(\frac{k}{k_\ast}\right)^{\Delta_2-1}, 
\end{align}
where $A_i = (\epsilon_{i,2}/2 + \Delta_i r_i)$ contains the second slow-roll parameter due to the derivative $y_i^\prime/y_i = \mathcal{H}_i (1+\epsilon_{i,2}/2)$. 

If Phase 1 is an exact de Sitter inflation with $\epsilon_1 =\delta_1 =0$, then $A_1 = \Delta_1 =0$ so that \eqref{boundary_constraint_2} gives $\mathcal{C}_{20} \sim (k/k_\ast)^{\Delta_1 - \Delta_2 +2}$ for $k/k_\ast \sim \mathcal{O}(1)$. Therefore the late-time curvature perturbation exhibits an unusual scaling as
\begin{align}
k^{3/2}\mathcal{R}_2(\eta) \rightarrow \mathcal{C}_{20} (-k\eta)^{\Delta_2} \sim k^{\Delta_1 +2}, 
\end{align}
for a certain range of $k$. This is nothing but a nearly $n_s -1 =4$ growth of the power spectrum ($P_{\mathcal{R}} \sim k^{2\Delta_1 + 4}$) when $\Delta_1 \ll 1$.
Note that for general cases with $\Delta_1 \neq 0$, \eqref{boundary_constraint_2} reaches to the same conclusion $\mathcal{C}_{20} \sim (k/k_\ast)^{\Delta_1 - \Delta_2}$ as that of \eqref{boundary_constraint_1}.
If Phase 2 approaches to the slow-roll limit with $A_2 \rightarrow 0$ and $ \Delta_2 \rightarrow 0$, then one will find that $\mathcal{C}_{2n} \rightarrow \mathcal{C}_{1n}$ and the first constraint \eqref{boundary_constraint_1} converges to be the same as \eqref{boundary_constraint_2}. 

\subsection{Bulk solutions}\label{Sec. bulk12}
We now perform a detailed calculation to support the boundary arguments in the previous section.
Taking $\mathcal{H}_i = k/\tau_i$, the equation of motion \eqref{eq:Mukhanov_sasaki} shares a similar structure as those of \eqref{eom:MS_1} and \eqref{eom:MS_2} so that the solution of $v_k$ in each phase takes the form of   
\begin{align}
v_1 &= c_1 \sqrt{\tau_1} H^{(1)}_{\nu_1} \left(\tau_1\right),  &&\quad \nu_1= 3/2 + \mathcal{O}(\epsilon), \\
v_2 &= c_2^{(1)} \sqrt{\tau_2} H^{(1)}_{\nu_2} \left(\tau_2 \right) + c_2^{(2)} \sqrt{\tau_2} H^{(2)}_{\nu_2} \left(\tau_2 \right) ,  &&\quad \nu_2= \vert \delta_2 + 3/2\vert,  
\end{align}
where $\nu_2 =  -\delta_2 - 3/2$ for $\delta_2 < -3/2$ and $c_1$ is given by \eqref{c_1}.
Matching the mode functions at $z = z_\ast$, $c_2^{(1)}$ and $c_2^{(2)}$ are solved by the Cauchy boundary conditions as
\begin{align}
\label{sol:c21}
c_2^{(1)} = \frac{i\pi}{4} c_1 &
\left[-\tau_\ast H^{(1)}_{\nu_1 -1}\left(\tau_\ast\right) H^{(2)}_{\nu_2}(\tau_\ast)  \right.\\\nonumber
&\;+ \left. H^{(1)}_{\nu_1}\left(\tau_\ast\right) \left(\tau_\ast H^{(2)}_{\nu_2-1}(\tau_\ast) +(A_2-A_1) H^{(2)}_{\nu_2}(\tau_\ast) \right) \right],
 \\\label{sol:c22}
c_2^{(2)} = \frac{i\pi}{4} c_1  &
\left[ \tau_\ast H^{(1)}_{\nu_1 -1}\left(\tau_\ast\right) H_{\nu_2}^{(1)}(\tau_\ast)  \right. \\\nonumber
&\;+ \left. H^{(1)}_{\nu_1}\left(\tau_\ast\right) \left( -\tau_\ast H^{(1)}_{\nu_2-1}(\tau_\ast) +(A_1-A_2) H^{(1)}_{\nu_2}(\tau_\ast) \right) \right],
\end{align}
with $A_i = (\epsilon_{i,2}/2 + \Delta_i r_i)$ for $i = 1,2$, where $\epsilon_1=\delta_1 =\epsilon_2 =0$ gives $A_1 = 0$ and $A_2 = 3 + 2\delta_2$.

\begin{figure}
	\begin{center}
		\includegraphics[width=9cm]{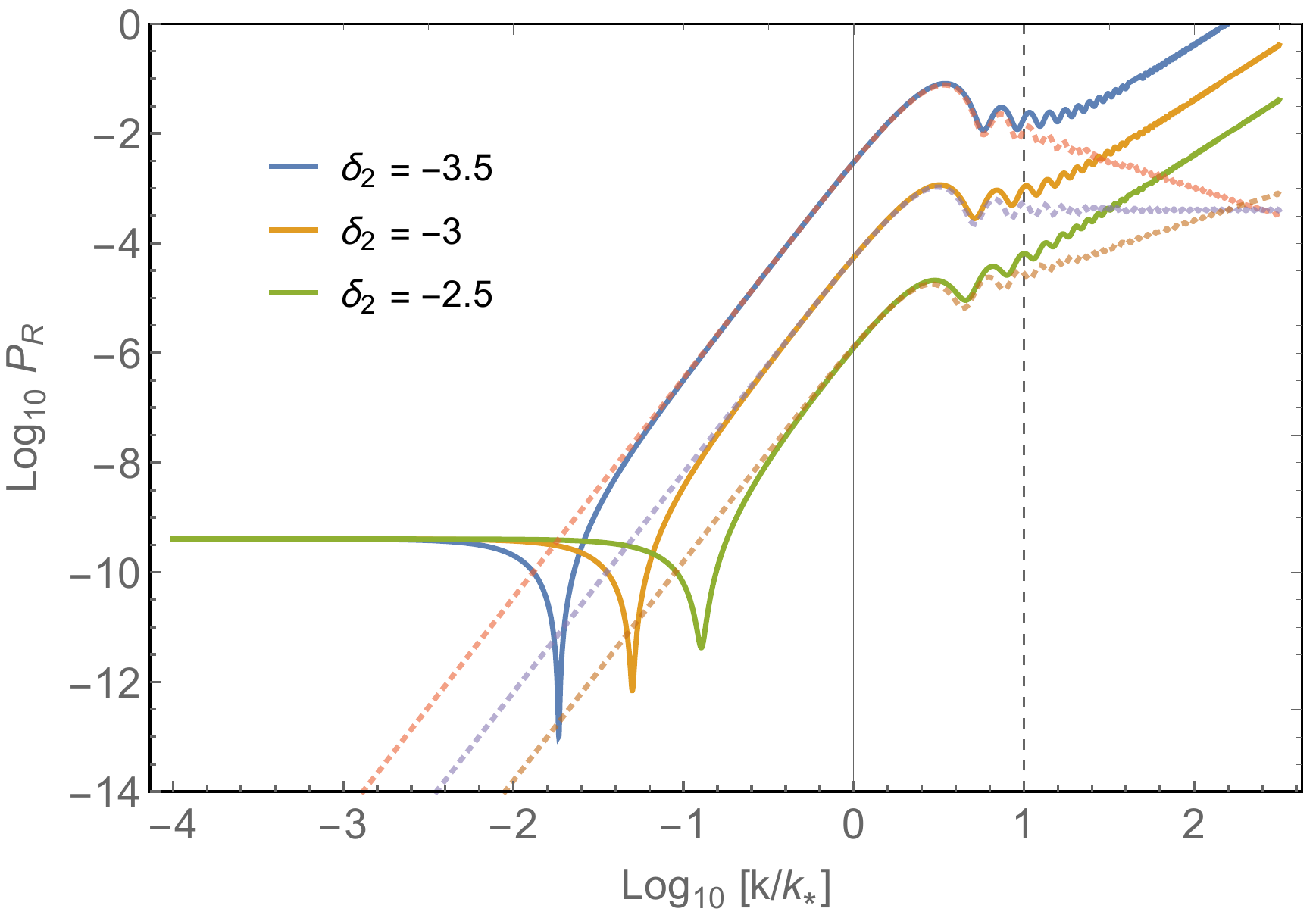}
	\end{center}
	\caption{The power spectrum $P_{\mathcal{R}}$ from the slow-roll (Phase 1) to the negative constant rate (Phase 2) given by \eqref{P_R_full} with $H_k = 10^{-5}$ in Planck unit. The dotted lines are IR expansions given by \eqref{P_R_IR}. The dashed-vertical line is $k = k_{\ast\ast}$. \label{fig:spectrum_phase2}}
\end{figure}

The power spectrum of $\mathcal{R}$ during Phase 2, $\eta_\ast < \eta \leq \eta_{\ast\ast}$, is given by
\begin{align}
P_{\mathcal{R}}(k; \eta) = \frac{k^3}{2\pi^2} \frac{1}{y^2(\eta)} \left\vert v_k(\eta)\right\vert^2 
= \frac{k^3}{2\pi^2} \frac{1}{a^2(\eta)\epsilon_2(\eta)} \left\vert v_2(\eta)\right\vert^2,
\end{align}
where $\epsilon_2(\eta) = \epsilon_\ast (\eta/\eta_\ast)^{-2\delta_2}$ and $\epsilon_\ast = \epsilon_1$ is a constant for the slow-roll phase probed on CMB scales. Denoting that $a_{\ast\ast} \equiv a(\eta_{\ast\ast})$, we have $a_\ast/a_{\ast\ast} = k_\ast/k_{\ast\ast}$, and the power spectrum at $\eta = \eta_{\ast\ast}$ reads
\begin{align} \label{P_R_full}
P_{\mathcal{R}}(k; \eta_{\ast\ast}) &= \frac{k^3}{2\pi^2} \frac{1}{a^2_k \epsilon_\ast} 
\left(\frac{k_\ast}{k_{\ast\ast}}\right)^{2\delta_2} \left(\frac{k}{k_{\ast\ast}}\right)^{2}\left\vert v_2(\eta_{\ast\ast})\right\vert^2
\nonumber\\
&= \frac{k H_k^2}{2\pi^2} \frac{1}{\epsilon_\ast} \left(\frac{k_\ast}{k_{\ast\ast}}\right)^{2+ 2\delta_2} \left(\frac{k}{k_{\ast}}\right)^{2}\left\vert v_2(\eta_{\ast\ast})\right\vert^2,
\end{align}
with $a_k = k/H_k$ and $k = \mathcal{H}(\eta_k) = -1/\eta_k$. This expression \eqref{P_R_full} reproduces the numerical result of the theoretical model in \cite{Cicoli:2018asa} with the analytical structure identified in \cite{Liu:2020oqe}.

To see the analytical structure of \eqref{P_R_full} more explicitly, let us focus on superhorizon modes, $k \ll k_{\ast\ast}$, by expanding the Hankel functions in $v_2$ to find that
\begin{align}\label{P_R_IR}
P_{\mathcal{R}}^{\rm IR}(k; \eta_{\ast\ast}) 
&= \frac{k H_k^2}{2\pi^2} \frac{1}{\epsilon_\ast}
\left(\frac{k_\ast}{k_{\ast\ast}}\right)^{2\delta_2} 
\left\vert c_1 f^2_{\mathcal{R}}(\tau_\ast)\right\vert^2  \left\vert \frac{i}{\pi}2^{\nu_2} \Gamma(\nu_2) \right\vert^2 \left(\frac{k}{k_{\ast\ast}}\right)^{3-2\nu_2}
\nonumber\\
& = \frac{ H_k^2}{8\pi} \frac{1}{\epsilon_\ast} \left\vert f^2_{\mathcal{R}}(\tau_\ast)\right\vert^2  \left\vert \frac{i}{\pi}2^{\nu_2} \Gamma(\nu_2) \right\vert^2 
\left(\frac{k_\ast}{k_{\ast\ast}}\right)^{6+4\delta_2} \tau_\ast^{6+2\delta_2},
\end{align}
where in terms of $\tau_\ast = -z_\ast = k/k_\ast$ we can transfer the ratio to $k/k_{\ast\ast} = \tau_\ast k_\ast/k_{\ast\ast}$.
One can see that $(k_\ast/k_{\ast\ast})^{6+4\delta_2} = e^{-(6+4\delta_2)\Delta N}$ is the factor of the enhancement due to a negative constant rate $\delta_2 < -3/2$ for a period of $\Delta N$ $e$-folding numbers \cite{Liu:2020oqe}.

The $k$-dependence of the spectrum is led by functions of $\tau_\ast$ 
where we have defined
\begin{align}\label{def:f2}
f^2_{\mathcal{R}}(\tau_\ast) &\equiv \frac{c_2^{(2)}- c_2^{(1)} }{c_1} \\\nonumber
&= \frac{i \pi}{2}  \tau_\ast 
\left[ 
J_{\nu_2}(\tau_\ast) H_{1/2}^{(1)}\left(\tau_\ast\right) + H_{3/2}^{(1)} \left(\tau_\ast\right) 
\left( \frac{-3-2\delta_2}{\tau_\ast}  J_{\nu_2} (\tau_\ast) -J_{\nu_2 -1} (\tau_\ast)  \right)
\right].
\end{align}
In the limit of $\tau_\ast \ll 1$, the leading terms of $f_{\mathcal{R}}$ are
\begin{align}
f^2_{\mathcal{R}}(\tau_\ast)  =-\sqrt{\frac{\pi}{2^3}} \left(\frac{\tau_\ast}{2}\right)^{\nu_2} 
\left[\frac{3+2\delta_2 +2\nu_2}{\Gamma(1+\nu_2)}\tau_\ast^{-3/2} +\frac{\delta_2}{2}\frac{7+2\delta_2 +2\nu_2}{\Gamma(2+\nu_2)} \tau_\ast^{1/2} + \cdots\right],
\end{align}
where the first term in proportion to $\tau_\ast^{\nu_2-3/2}$ vanishes accidentally since $\nu_2 = -\delta_2 -3/2$ for $\delta_2 < -3/2$. Therefore $f^2_{\mathcal{R}} \sim \tau_\ast^{\nu_2 + 1/2}$ for $k < k_\ast$ and the power spectrum $P_{\mathcal{R}}^{\rm IR} \sim \tau_\ast^4$  shows a $k^4$ enhancement for $k_{\rm min} < k < k_\ast$ independent of the value of $\delta_2$. The minimal scale of growth $ k_{\rm min}/ k_\ast = (k_\ast/k_{\ast\ast})^{\nu_2}$ is estimated by the condition $P_\mathcal{R}(k = k_\ast) = (k_\ast/k_{\ast\ast})^{6+2\delta_2} P_\mathcal{R}(k \ll k_\ast)$  \cite{Liu:2020oqe}. For $k < k_{\rm min}$, the constant modes dominate $P_{\mathcal{R}}$, thus reproducing the standard $k$-invariant result on very large scales.

On the other hand, in the limit of $\tau_\ast \gg 1$, $f^2_{\mathcal{R}} \rightarrow e^{i \pi(1+2\nu_2)/4}$ becomes a purely oscillatory function of $k$ so that the power-law scaling $P_{\mathcal{R}}\sim \tau_\ast^{2\Delta_2}=\tau_\ast^{6+2\delta_2}$ for $k > k_\ast$.
This describes the power spectrum for $k$ modes exit the horizon well in side Phase 2.
The maximal of $P_{\mathcal{R}}$ appears at $k = k_\ast$ if $\delta_2 < -3$ and at $k = k_{\ast\ast}$ if $-3 < \delta_2 < -3/2$.
In Figure~\ref{fig:spectrum_phase2} we plot the full expression \eqref{P_R_full} and the IR approximation \eqref{P_R_IR} of the power spectrum solved in this section with different choices of $\delta_2 < -3/2$. In the limit of $\tau_\ast \rightarrow \infty$, $v_2(\eta_{\ast\ast})$ has no additional power-law dependence on $k$ (see the discussion of \textit{the large $k$ limit} in Section~\ref{Sec. N_stage} for more details) so that $P_{\mathcal{R}} \sim \tau_\ast^2$, while $P_{\mathcal{R}}^{\rm IR} \sim \tau_\ast^{6+2\delta_2}$ holds for the extrapolated regime beyond $k > k_{\ast\ast}$. 
For the exact USR case with $\delta_2 = -3$, one finds that $P_{\mathcal{R}}^{\rm IR} \sim \tau_\ast^{0}$ stands for a scale-invariant spectrum in the large $k$ limit.  
The limitation of the continuous scaling of $P_{\mathcal{R}}$ after the end of Phase 2 is discussed in Section~\ref{Sec. reduce_R}.

\section{Continuous decay of power spectrum}\label{Sec. reduce_R}
The duration of Phase 2, namely $\Delta N = \ln (k_{\ast\ast}/k_\ast)$, with a negative constant rate, $\delta_2 < -3/2$, is the key factor that determines the magnitude of enhancement to the power spectrum at the beginning of Phase 2: $P_{\mathcal{R}}(k = k_\ast)/P_{\mathcal{R}}(k \ll k_\ast) \sim(k_\ast/k_{\ast\ast})^{6+4\delta_2} = e^{-(6+4\delta_2)\Delta N}$. This means that the ending time of Phase 2, namely $\eta =\eta_{\ast\ast} = -1/k_{\ast\ast}$, cannot be arbitrarily extended (to some number close to zero) or otherwise the power spectrum becomes arbitrarily large. Thus a realistic power spectrum on the boundary surface at the end of inflation must include also the solutions of mode functions in the post negative-constant-rate phase (Phase 3). In particular, to realize a spiky power spectrum for PBH formation around a certain mass scale, Phase 3 shall address how the curvature perturbation can be successfully reduced from the enhancement after Phase 2. 

\subsection{Boundary arguments for continuous scaling}\label{Sec. boundary23}
With a careful look at the numerical results of \cite{Cicoli:2018asa,Liu:2020oqe,Cheng:2018qof}, one finds that $\delta_3$, the rate of rolling in the post negative-constant-rate phase, is still approximately a constant and it can have a positive value as large as $\mathcal{O}(1)$, which significantly violates the slow-roll conditions. Based on this observation we construct Phase 3 with a constant rate parameter $\delta_3 \geq 0$, 
and we expect that the late-time scaling of the curvature perturbation is again led by the conformal weight $\Delta_3$ with respect to the dilatation symmetry as
\begin{align}\label{def:late_time_R3}
k^{3/2}\mathcal{R}_3 \rightarrow \mathcal{C}_{30} (-k\eta)^{\Delta_3} + \mathcal{C}_{3 1} (-k\eta)^{\Delta_3 + 2} + \cdots,
\end{align}
where $\Delta_3 = 3/2-\nu_3 = - \delta_3$ can include the slow-roll case if $\delta_3 \ll 1$.

As argued in Section~\ref{Sec. boundary12}, the boundary condition $\mathcal{R}_2(\eta_{\ast\ast}) = \mathcal{R}_3(\eta_{\ast\ast})$ for $k\ll k_{\ast\ast}$ indicates $\mathcal{C}_{30} \sim  \mathcal{C}_{20}(k/k_{\ast\ast})^{\Delta_2 -\Delta_3}$ so that $k^{3/2}\mathcal{R}_3 \sim  \mathcal{C}_{20} k^{\Delta_2} $ shows a scaling independent of $\Delta_3$ in Phase 2. Note that $\mathcal{C}_{20} \propto c_1 f_\mathcal{R}^2$ given by \eqref{def:f2} becomes $k$-independent in the limit of $k \gg k_\ast$, and therefore $k^{3/2}\mathcal{R}_3 \sim  k^{\Delta_2}$ reproduces the continuous scaling $P_\mathcal{R}\sim k^{2\Delta_2} =k^{6+2\delta_2}$ in \cite{Byrnes:2018txb,Cheng:2018qof,Cicoli:2018asa,Liu:2020oqe} to Phase 3 where inflaton is no longer rolling with a negative constant rate.


The continuous $k$-scaling of the curvature perturbation from Phase 2 to Phase 3 breaks down when $k/k_{\ast\ast} \gg 1$. Since $\mathcal{R}_i(k\rightarrow \infty) \sim c_i^{(1,2)}e^{\pm i k\eta}/y_i$, the boundary condition $\mathcal{R}_1(\eta_{\ast}) = \mathcal{R}_2(\eta_{\ast})$ in the large $k$ limit indicates that the absolute value $\vert\mathcal{C}_{20}\vert$ is $k$-independent. Similarly, $\vert\mathcal{C}_{30}\vert$ is $k$-independent in the limit of $k\rightarrow \infty$ due to the boundary condition $\mathcal{R}_2(\eta_{\ast\ast}) = \mathcal{R}_3(\eta_{\ast\ast})$. As a result, the asymptotic momentum scaling of the  power spectrum reads $P_{\mathcal{R}} \sim k^{2\Delta_3} = k^{-2\delta_3}$ in the large $k$ limit.

\subsection{Unified bulk solutions}\label{Sec. bulk3}
We now derive the full analytic solutions to be observed on the final boundary at the end of Phase 3.
The mode function of the Mukhanov-Sasaki variable in Phase 3 takes the form of
\begin{align}
v_3 = c_3^{(1)} \sqrt{\tau_3} H^{(1)}_{\nu_3} \left(\tau_3 \right) + c_3^{(2)} \sqrt{\tau_3} H^{(2)}_{\nu_3} \left(\tau_3 \right) ,  &&\quad \nu_3=  \delta_3 + 3/2,  
\end{align}
where $\tau_3 = -k\eta$ and $\mathcal{H}_3 = k/\tau_3 = -1/\eta$ by taking $\epsilon_3 = 0$.

Let us define $k_{\rm end} = -1/\eta_{\rm end}$ as the scale at the end of Phase 3 and the duration of Phase 2 and 3 are fixed by the ratios $x_2 \equiv k_{\ast}/k_{\ast\ast}$, $x_3 \equiv k_{\ast\ast}/k_{\rm end}$ which are always smaller than unity. This gives $z_{\rm end} = -k\eta_{\rm end} = \tau_\ast x_2 x_3$, and the Phase 3 coefficients solved at the boundary surface $-\eta = k_{\ast\ast}$ are given by
\begin{align}
c_3^{(1)} =& \frac{i\pi}{4} c_1 \tau_\ast x_2 
\left[  \left(c_2^{(1)} H_{\nu_2}^{(1)}(\tau_\ast x_2  ) 
+c_2^{(2)} H_{\nu_2}^{(2)} (\tau_\ast x_2  ) \right) H^{(2)}_{\nu_3 -1}(\tau_\ast x_2  ) \right.
 \\\nonumber
&\;+ \left. \left(c_2^{(1)}H^{(1)}_{\nu_2+1} (\tau_\ast x_2 ) 
+ c_2^{(2)}  H^{(2)}_{\nu_2+1} (\tau_\ast\tau_{\ast\ast} ) \right) H_{\nu_3}^{(2)} (\tau_\ast x_2 ) \right], \\
c_3^{(2)} =& -\frac{i\pi}{4} c_1 \tau_\ast x_2 
\left[  \left(c_2^{(1)} H_{\nu_2}^{(1)}(\tau_\ast x_2 ) 
+c_2^{(2)} H_{\nu_2}^{(2)} (\tau_\ast x_2  ) \right) H^{(1)}_{\nu_3 -1}(\tau_\ast x_2  ) \right.
\\\nonumber
&\;+ \left. \left(c_2^{(1)}H^{(1)}_{\nu_2+1} (\tau_\ast x_2  ) 
+ c_2^{(2)}  H^{(2)}_{\nu_2+1} (\tau_\ast x_2 ) \right) H_{\nu_3}^{(1)} (\tau_\ast x_2  ) \right]. 
\end{align}
Note that $c_2^{(1,2)}$ solved by \eqref{sol:c21} and \eqref{sol:c22} are scale-dependent functions of $\tau_\ast = k/k_\ast $.

Since $\delta_3 > 0$ implies an enhancement of $\epsilon_3$, we find that $\epsilon_{\rm end} = \epsilon(\eta_{\rm end}) = \epsilon_\ast x_2^{-2\delta_2}x_3^{-2\delta_3}$. The final power spectrum is then
\begin{align} \label{P_R_phase3_full}
P_{\mathcal{R}}(k; \eta_{\rm end}) &= \frac{k^3}{2\pi^2} \frac{1}{a^2_{\rm end} \epsilon_{\rm end}} 
\left\vert v_3(\eta_{\rm end})\right\vert^2
\nonumber\\
&= \frac{k H_k^2}{2\pi^2} \left(\tau_\ast x_2 x_3\right)^{2}  \frac{x_2^{2\delta_2} x_3^{2\delta_3}}{\epsilon_\ast}
\left\vert v_3(\eta_{\rm end})\right\vert^2, 
\end{align}
and the expansion of $v_3$ for $k/k_{\rm end} \ll 1$ gives 
\begin{align}\label{P_R_phase3_IR}
P_{\mathcal{R}}^{\rm IR}(k; \eta_{\rm end}) 
&= \frac{ H_k^2}{8\pi} \frac{x_2^{2\delta_2} x_3^{2\delta_3}}{\epsilon_\ast} 
\left\vert f_{\mathcal{R}}^3(\tau_\ast, x_2)\right\vert^2  \left\vert \frac{i}{\pi}2^{\nu_3} \Gamma(\nu_3) \right\vert^2 
\left(\tau_\ast x_2 x_3\right)^{3-2\nu_3}\\
&= \frac{ H_k^2}{8\pi} \frac{x_2^{2\delta_2} }{\epsilon_\ast} 
\left\vert f_{\mathcal{R}}^3(\tau_\ast, x_2)\right\vert^2  \left\vert \frac{i}{\pi}2^{3/2 +\delta_3} \Gamma(3/2 +\delta_3) \right\vert^2 
\left(\tau_\ast x_2 \right)^{-2\delta_3},
\end{align}
where we have defined $f_{\mathcal{R}}^3 \equiv (c_3^{(2)} - c_3^{(1)})/c_1$.
We remark that the general expression in terms of $\nu_2$ and $\nu_3$ applies to arbitrary values of $\delta_2$ and $\delta_3$, while the specific expression using $\nu_2 = -3/2 -\delta_2$ and $\nu_3 = 3/2 +\delta_3$ is for $\delta_2 < -3/2$ and $\delta_3 \geq 0$.

\begin{figure}
	\begin{center}
		\includegraphics[width=7cm]{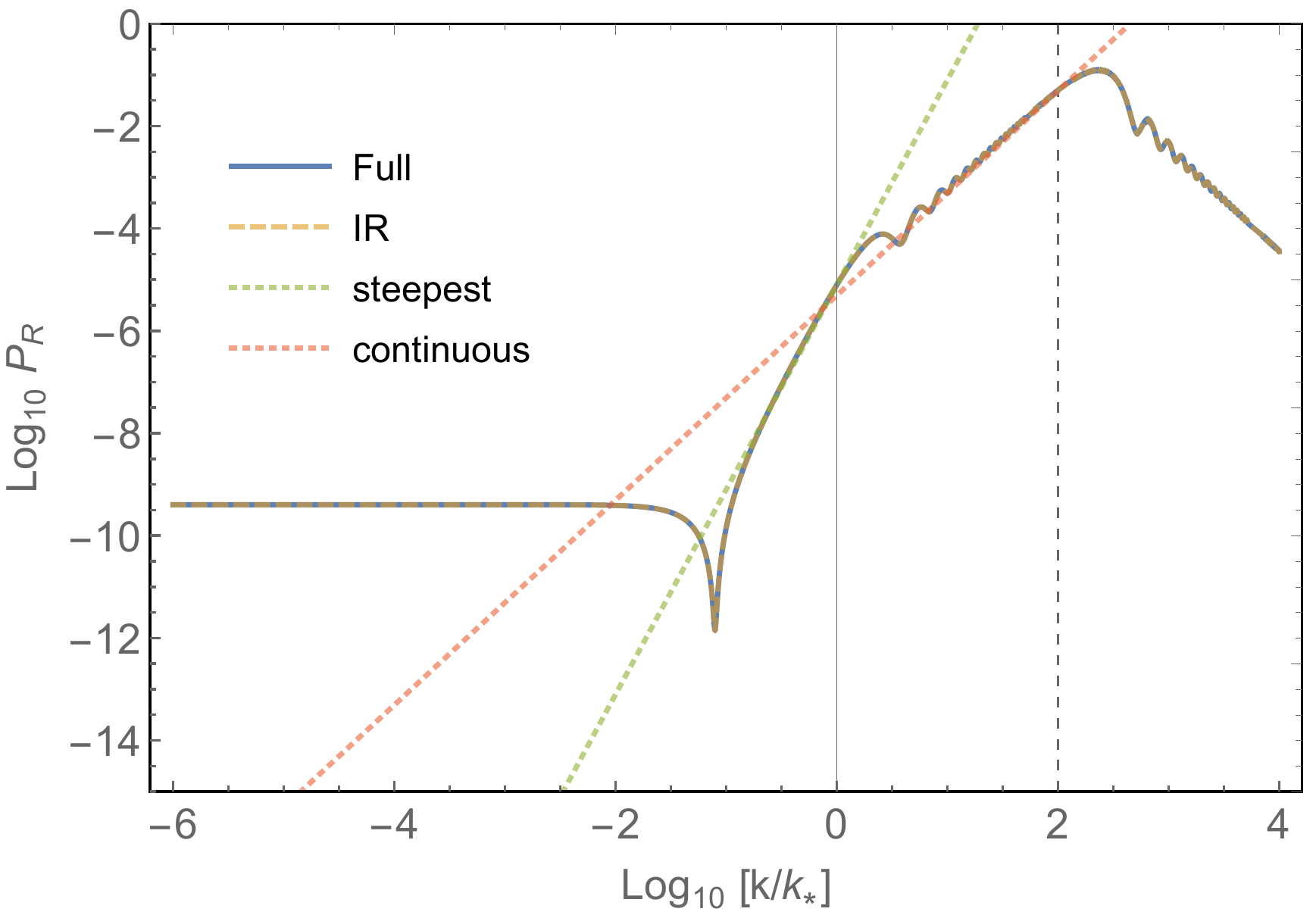}
		\hfill
		\includegraphics[width=7cm]{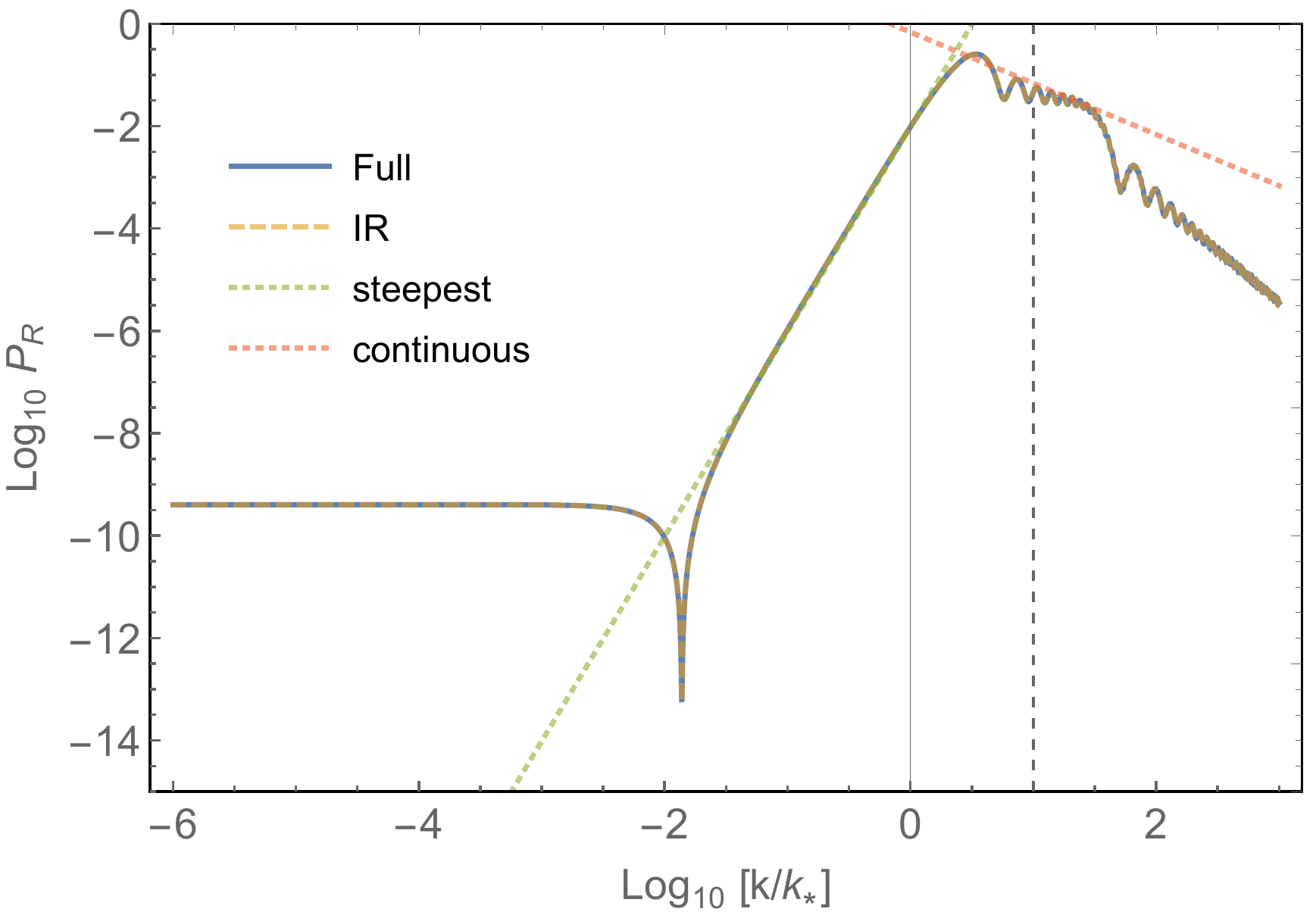}
	\end{center}
	\caption{The power spectrum $P_{\mathcal{R}}$ given by \eqref{P_R_phase3_full} with $H_k = 10^{-5}$ in Planck unit from the phases of slow-roll (Phase 1) to negative-constant-rate (Phase 2) with $\delta_2 =-2$ (left panel) and $\delta_2 = -3.5$ (right panel). The dashed-vertical line is $k = k_{\ast\ast}$, and for $k >  k_{\ast\ast}$ a positive-constant-rate $\delta_3 = 1$ is used in Phase 3. The late-time (IR) expansions of the power spectrum \eqref{P_R_phase3_IR} in terms of the dashed lines are in well agreement with the full expression \eqref{P_R_phase3_full} in solid lines. \label{fig:spectrum_phase3_p1}}
\end{figure}

\begin{figure}
	\begin{center}
		\includegraphics[width=7cm]{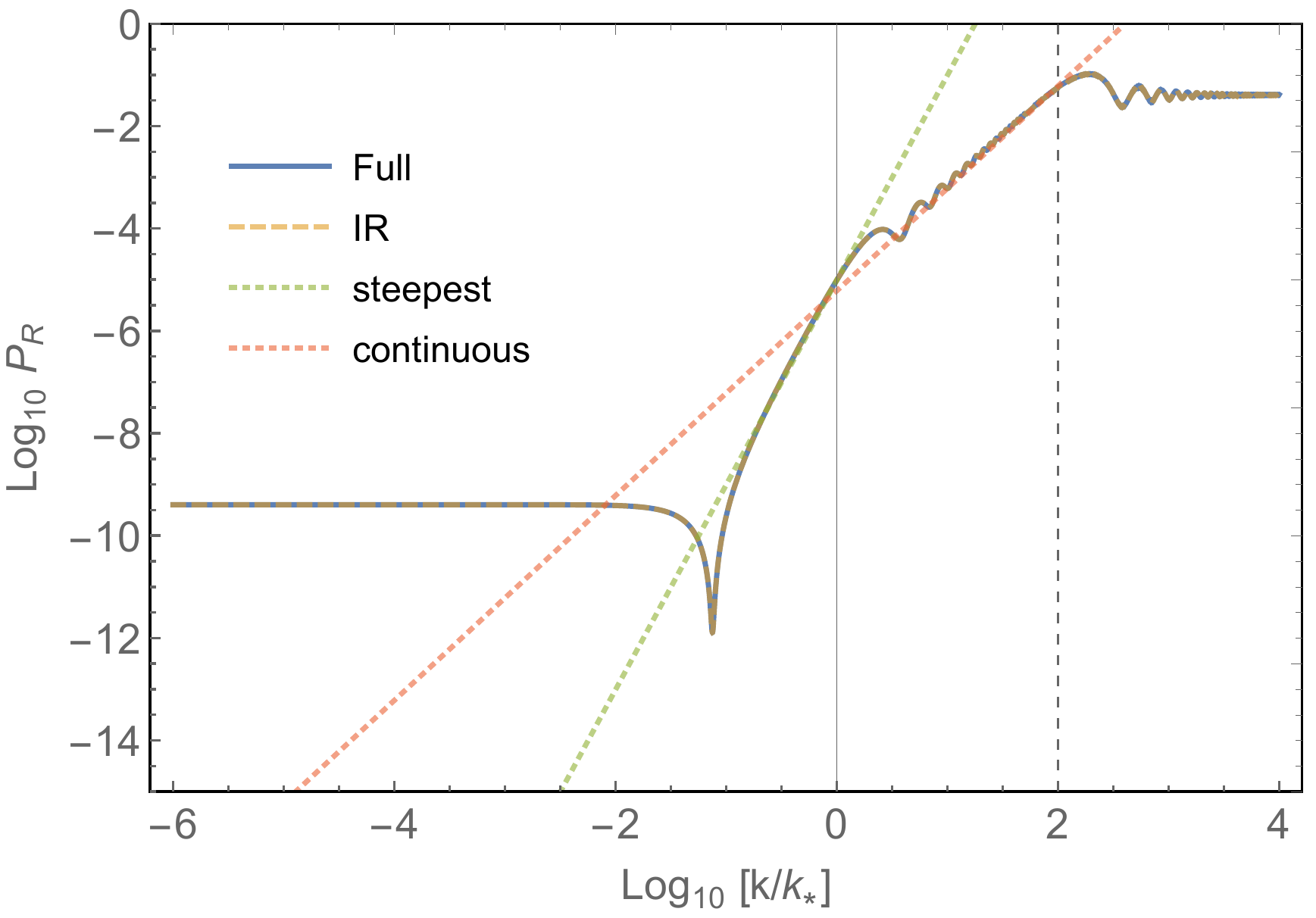}
		\hfill
		\includegraphics[width=7cm]{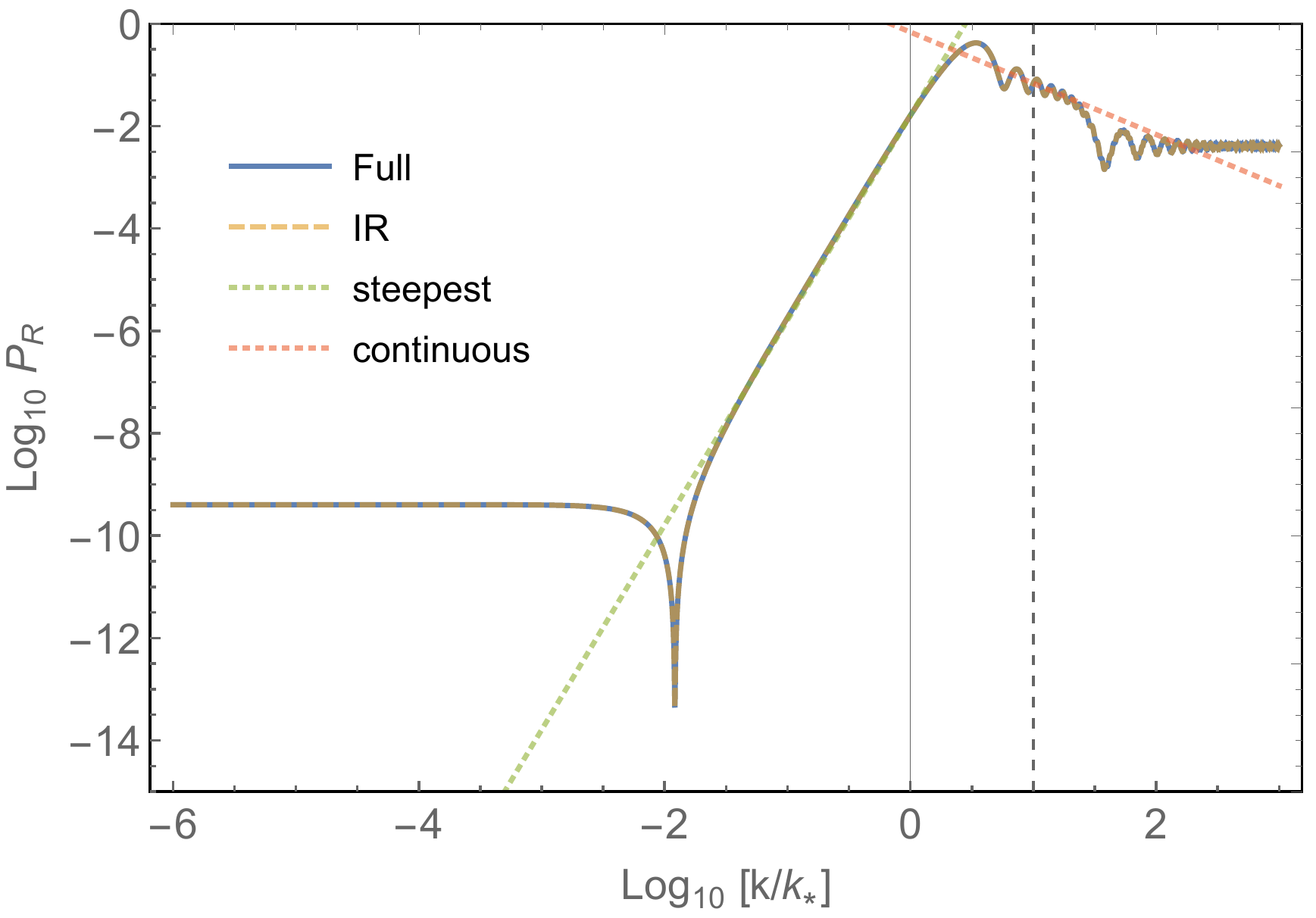}
	\end{center}
	\caption{The power spectrum $P_{\mathcal{R}}$ given by \eqref{P_R_phase3_full} with $H_k = 10^{-5}$ in Planck unit from the phases of slow-roll (Phase 1) to negative-constant-rate (Phase 2) with $\delta_2 =-2$ (left panel) and $\delta_2 = -3.5$ (right panel). The dashed-vertical line is $k = k_{\ast\ast}$, and for $k >  k_{\ast\ast}$ a secondary slow-roll phase $\delta_3 = 0$ is used in Phase 3. The late-time (IR) expansions of the power spectrum \eqref{P_R_phase3_IR} in terms of the dashed lines are in well agreement with the full expression \eqref{P_R_phase3_full} in solid lines. \label{fig:spectrum_phase3_SR}}
\end{figure}

In Figure~\ref{fig:spectrum_phase3_p1}, we plot examples with a positive constant rate $\delta_3 > 0$ in Phase 3, and in Figure~\ref{fig:spectrum_phase3_SR}, we plot examples with $\delta_3 = 0$, where Phase 3 is effectively a slow-roll inflation.
\footnote{Single-field models of inflation for PBH formation ending with an effective slow-roll phase can be found in \cite{Byrnes:2018txb,Biagetti:2018pjj,Motohashi:2019rhu,Ragavendra:2020sop}.}
In the left (right) panel of Figures~\ref{fig:spectrum_phase3_p1} and \ref{fig:spectrum_phase3_SR}, we use $-3 < \delta_2 < -3/2$ (and $\delta_2 < -3$) so that the Phase 2 spectrum $P_\mathcal{R} \sim k^{2\Delta_2} = k^{6 +2\delta_2}$ is a growing (decaying) phase, respectively.
One can see that the steepest growth $P_\mathcal{R} \sim k^4$, outlined by the ``steepest'' dotted lines, starts from $k_{\rm min} \approx k_\ast x_1^{\nu_2}$ in Phase 1 and continues to the beginning in Phase 2. Similarly, a continuous scaling of $P_\mathcal{R} \sim k^{2\Delta_2}$, outlined by the ``continuous'' dotted lines, occurs from Phase 2 to the beginning of Phase 3 for arbitrary choices of $\delta_3$. All examples indicate that the continuous scaling from Phase 2 breaks down in the limit of $k/k_{\ast\ast} \gg 1$ and the asymptotic behavior is again led by the conformal weight of Phase 3 as $P_\mathcal{R} \sim k^{2\Delta_3}$.

\subsection{PBH scenarios}\label{Sec. PBH scenario}
So far we have derived analytic power spectra from instantaneous transitions of phases with different constant rate-of-rollings. 
To obtain the unified formula \eqref{P_R_phase3_IR}, there is no assumption applied to the asymptotic scaling dimension towards the end of inflation (here we identify as $\Delta_3$) so that $\delta_3$ in general can be a free parameter that does not necessarily relate to the negative constant rate $\delta_2$ in Phase 2.
This implies that PBH scenarios with a continuous scaling to the end of inflation should involve with non-trivial constraints.
To see this, let us divide the existing models into two classes with respect to the scaling behavior of the power spectrum on scales much smaller than the (steepest) growing phase led by a negative $\delta < -3$. We refer scenarios with an approximately constant scaling dimension $\vert\Delta(N \rightarrow N_{\rm end})\vert \sim \mathcal{O}(1)$
that significantly violates the standard slow-roll conditions as the first class    \cite{Atal:2018neu,Byrnes:2018txb,Cicoli:2018asa,Garcia-Bellido:2017mdw,Liu:2020oqe,Cheng:2018qof,Biagetti:2018pjj,Germani:2017bcs},
and those with $\vert\Delta (N \rightarrow N_{\rm end})\vert \ll 1$ being essentially a secondary slow-roll phase as the second class \cite{Byrnes:2018txb,Biagetti:2018pjj,Motohashi:2019rhu,Ragavendra:2020sop}.   

To demonstrate the PBH scenarios in both secondary slow-rolling (SSR) and slow-roll-violating (SRV) classes, we adopt the parametrization of the inflaton potential by a power-law series \cite{Cheng:2018qof} as
\begin{align}
V(\phi) = V_0 \left[\alpha_0 + \alpha_1 \frac{\phi}{\Lambda} + \frac{\alpha_2}{2}\left(\frac{\phi}{\Lambda}\right)^2 +
\frac{\alpha_3}{3!}\left(\frac{\phi}{\Lambda}\right)^3+\frac{\alpha_4}{4!}\left(\frac{\phi}{\Lambda}\right)^4 +
\frac{\alpha_5}{5!}\left(\frac{\phi}{\Lambda}\right)^5  \right],
\end{align}  
where $\Lambda$ measures the vacuum expectation value of $\phi$. 
The power spectrum is solved by the numerical evaluation of $\epsilon(N)$ with respect to the $e$-folding number $N = \ln a$ as $P_\mathcal{R}(N) = H_\ast^2/(2\pi^2 M_{\rm pl}^2 \epsilon(N))$, where $H_\ast$ is the initial value of the Hubble parameter.

\begin{figure}
	\begin{center}
		\includegraphics[width=7cm]{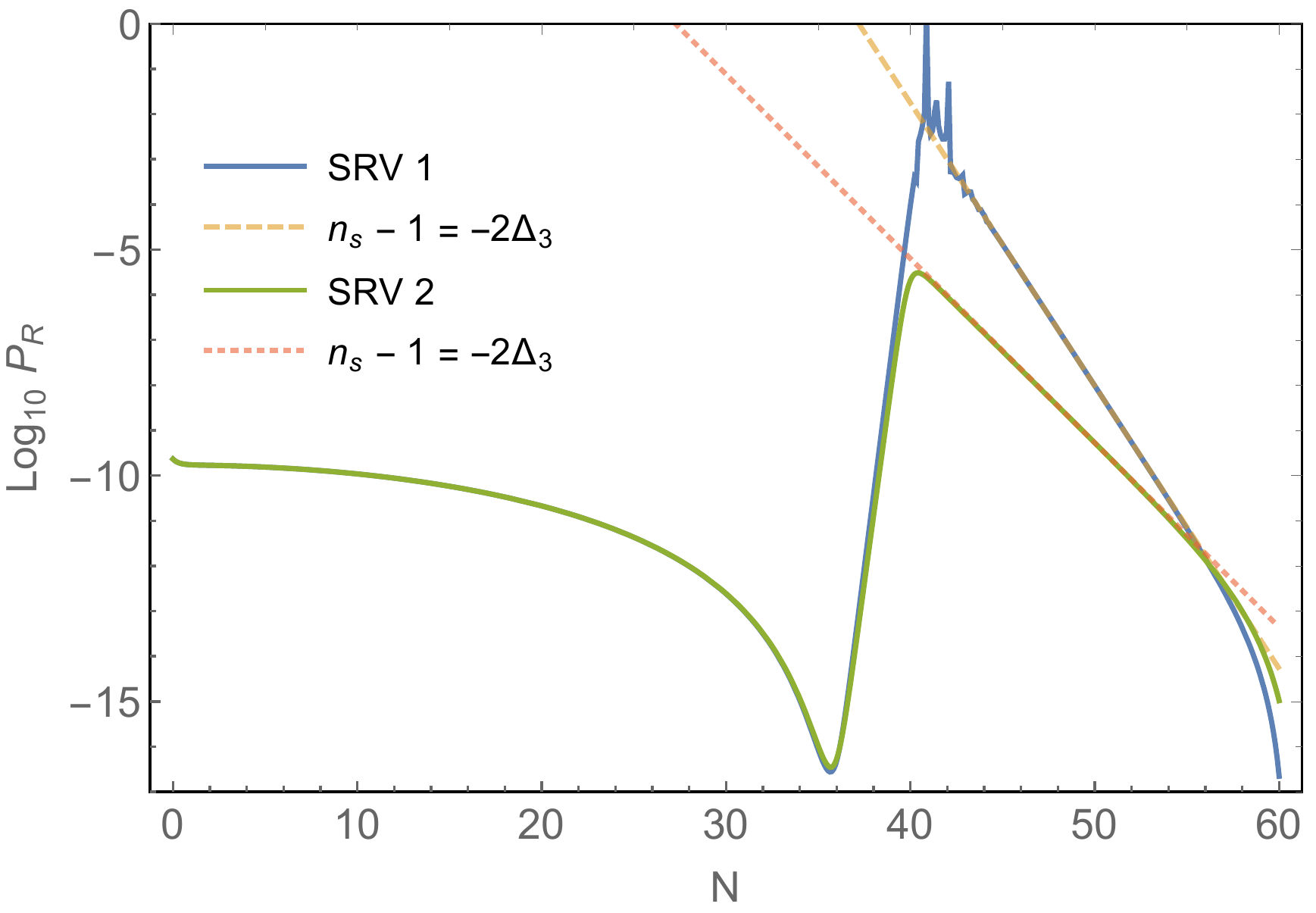}
		\hfill
		\includegraphics[width=7.5cm]{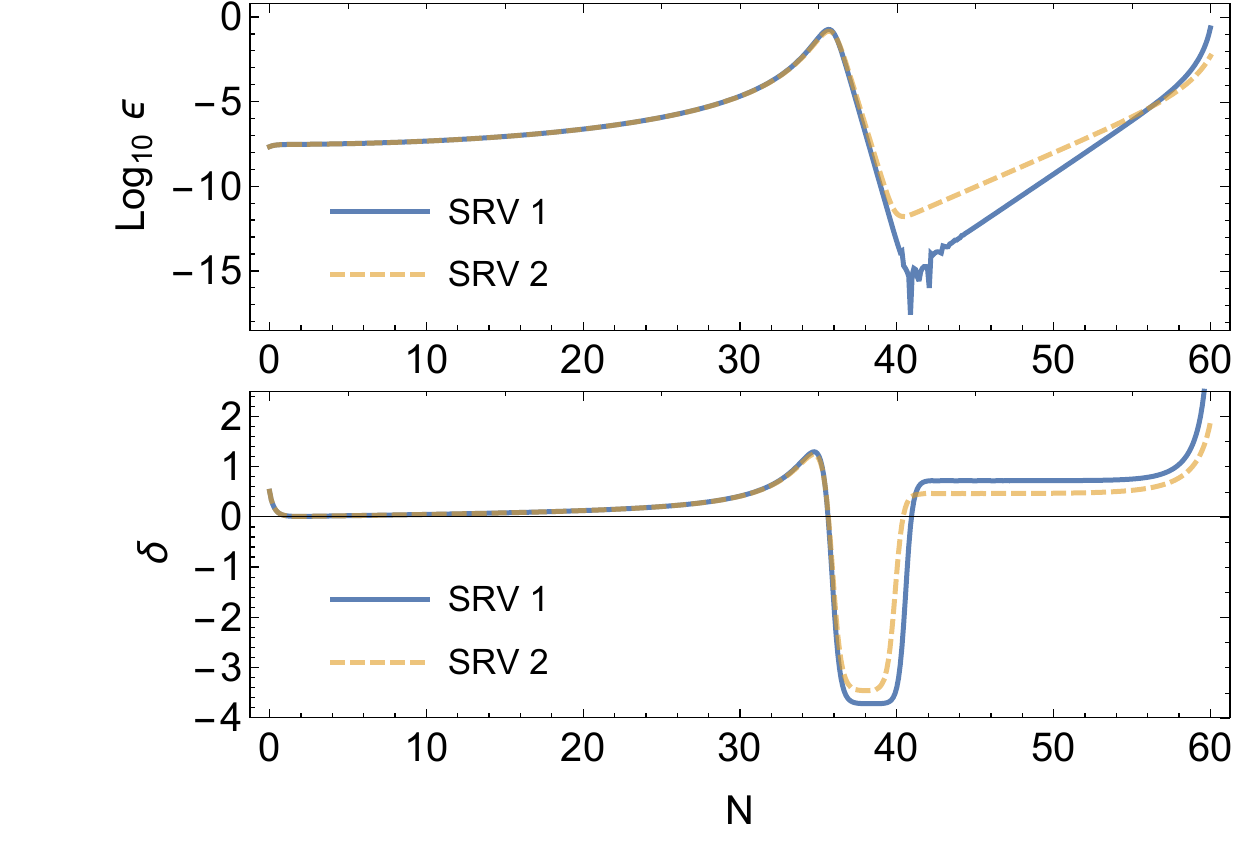}
	\end{center}
	\caption{The power spectrum $P_{\mathcal{R}}$ of the slow-roll-violating (SRV) class with the transition from $\delta < 0$ to $\delta > 0$. $H_\ast = 10^{-8} M_{\rm pl}$ is used in this plot. The evolution of $\delta$ from negative to positive values is constrained by the continuity of the scaling dimension (conformal weight) $\Delta$. The dotted lines are fitted power spectra with the spectral index given by $n_s - 1 = -2\Delta_3$, where $\Delta_3 = \Delta(N =50)$ is used. \label{fig:SRV}}
\end{figure}

For SRV scenarios with the slow-roll violation in the decaying phase of the power spectrum till the end of inflation ($\Delta > 0$ for $N \rightarrow N_{\rm end}$), we use $\Lambda = 0.3 M_{\rm pl}$, $V_0 = 3M_{\rm pl}^2 H_\ast^2$, $\alpha_0 = 1$ and $\alpha_2 = 0$. This choice subjects to a kind of small field inflation potential and therefore the initial value is taken in the limit of $\phi/\Lambda \ll 1$. In Figure~\ref{fig:SRV}, we fix $\{\alpha_1, \alpha_3 \} =\{ -0.73\times 10^{-4}, -0.52\}$ and change $\{\alpha_4, \alpha_5 \}$ to obtain different values of the rolling rate $\delta (N)$ for the $e$-folding numbers $N \gtrsim 36$. For SRV 1 (SRV 2) in Figure~\ref{fig:SRV}, $\alpha_4 = 0.95$ ($\alpha_4 = 1$) is used and $\alpha_5 < 0$ must be accurately tuned such that the inflaton has just enough momentum to roll away from the inflection point or otherwise a period of stable $\delta < 0$ cannot be realized. Such a fine-tuning condition ensures $\Delta_2 = 3 +\delta_2 \approx \Delta_3 = -\delta_3$, so that the scaling dimension found in Phase 2 (see Section~\ref{Sec. analytic_s}) approximately continues to Phase 3. For the evolution of $\delta(N)$ with respect to $N$, we have $\{\delta(38), \delta(50)\} =\{ -3.718, 0.721\}$ for SRV 1 and $\{\delta(38), \delta(50)\} =\{ -3.458, 0.470\}$ for SRV 2, which imply nearly constant conformal weights:$\{ \Delta_2 , \Delta_3  \} = \{-0.718, -0.721\}$ for SRV 1 and $\{ \Delta_2 , \Delta_3  \} = \{-0.458, -0.470\}$ for SRV 2. 

\begin{figure}
	\begin{center}
		\includegraphics[width=7cm]{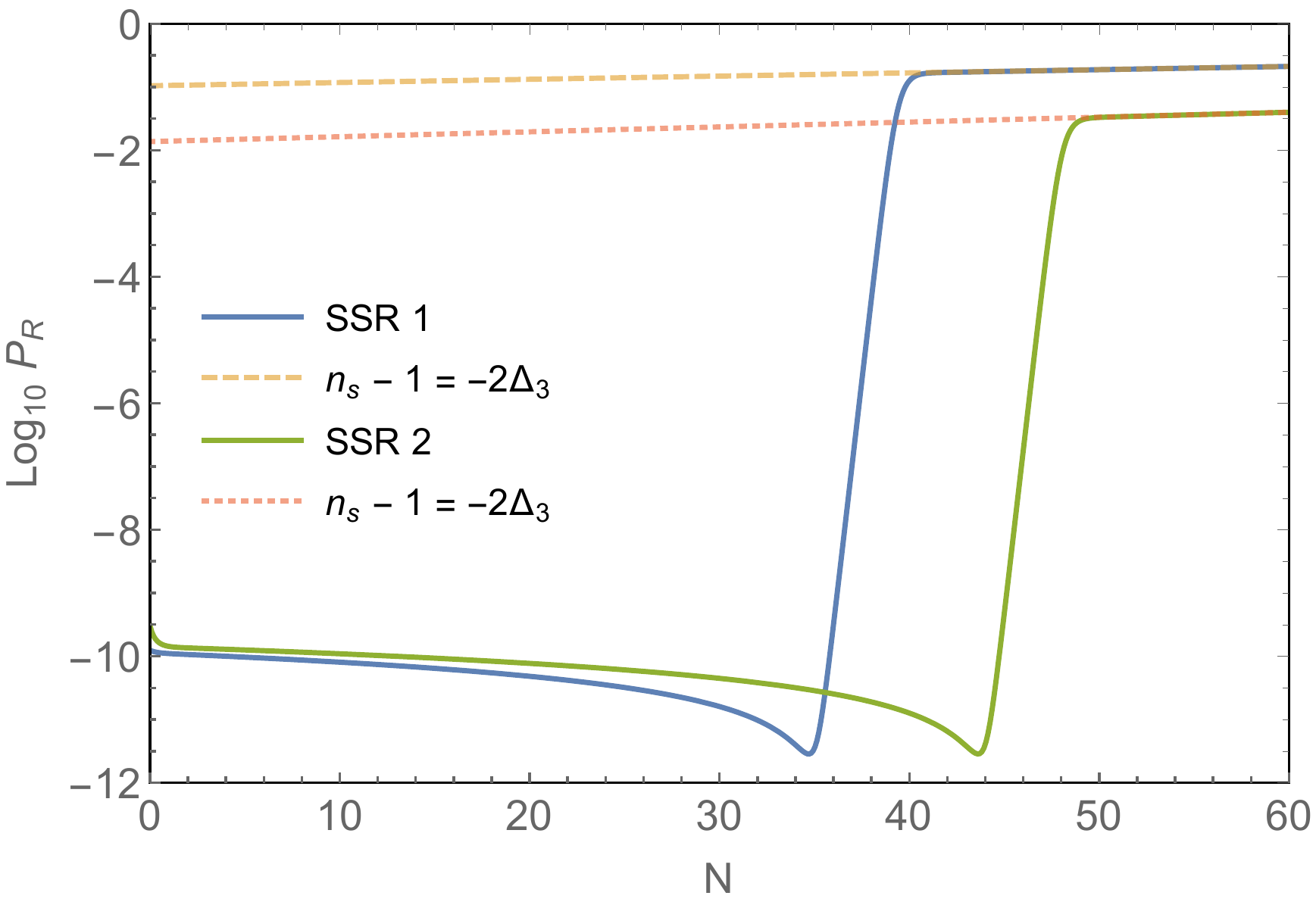}
		\hfill
		\includegraphics[width=7.5cm]{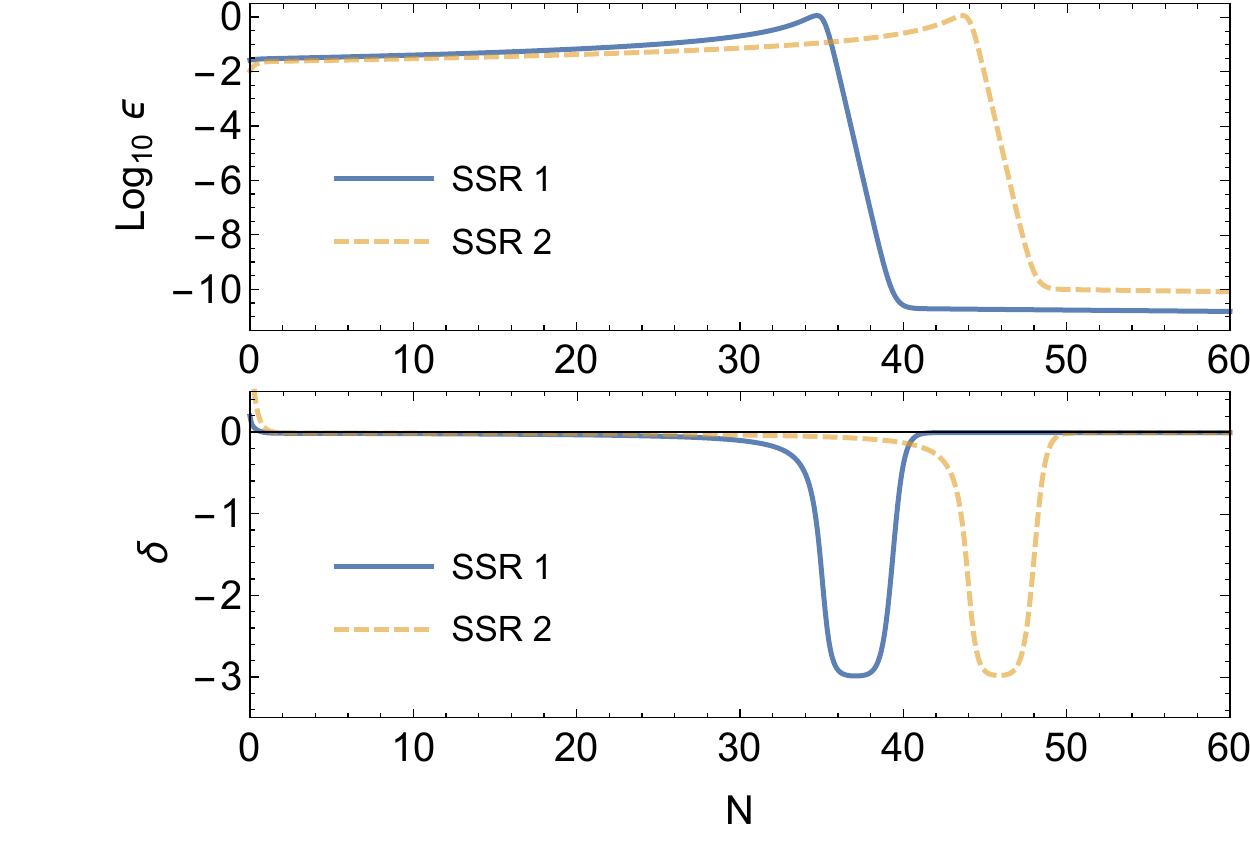}
	\end{center}
	\caption{The power spectrum $P_{\mathcal{R}}$ of the secondary slow-rolling (SSR) class with the transition from $\delta < 0$ to $\delta \approx 0$. $H_\ast =8 \times10^{-6} M_{\rm pl}$ is used in this plot. The dotted lines are fitted power spectra with the spectral index given by $n_s - 1 = -2\Delta_3$, where $\Delta_3 = \Delta(N =60)$ is used. \label{fig:SR}}
\end{figure}

For SSR scenarios where a secondary slow roll follows the growing phase led by a negative rate ($\Delta \approx 0$ for $N \rightarrow N_{\rm end}$), we use $\Lambda = M_{\rm pl}$ and $\alpha_0 =\alpha_1 =\alpha_5 =0$. This choice subjects to a specific potential form of the punctuated inflation \cite{Ragavendra:2020sop,Jain:2008dw,Allahverdi:2006we,Jain:2009pm}. In Figure~\ref{fig:SR}, we use $\{\alpha_2, \alpha_3, \alpha_4\} = \{ 1, -4/\phi_\ast, 6/\phi_\ast^2\}$ with $\phi_\ast = 1.983 M_{\rm pl}$, $V_0/(M_{\rm pl}^2 H_\ast^2) = 6.1\times 10^{-4}$ and the initial value $\phi_0 = 18 M_{\rm pl}$ for SSR 1. For SSR 2, we choose $\{\alpha_2, \alpha_3, \alpha_4\} = \{ 1, -4/\phi_\ast, 6/\phi_\ast^2\}$ with $\phi_\ast = 1.985 M_{\rm pl}$, $V_0/(M_{\rm pl}^2 H_\ast^2) = 3.9\times 10^{-4}$, and $\phi_0 = 20 M_{\rm pl}$. We find that $\{\delta(37),\delta(60)\} =\{  -2.987 , -5.86 \times 10^{-3}\}$ for SSR 1 and $\{\delta(46),\delta(60)\} =\{  -2.980 , -8.89 \times 10^{-3}\}$ for SSR 2. These results also manifest the approximately continuous scaling dimensions between the range of $\{ \Delta_2 , \Delta_3  \} = \{ \Delta(37), \Delta(60)\} = \{1.27 \times 10^{-2}, 5.86 \times 10^{-3}\}$ for SSR 1 or $\{ \Delta_2 , \Delta_3  \} = \{ \Delta(46), \Delta(60)\} = \{1.98 \times 10^{-2}, 8.89 \times 10^{-3}\}$ for SSR 2.

In summary, the asymptotic scaling dimension towards the end of inflation, $\Delta_3$, is in principle a free parameter governed by the inflaton potential. However, an inflation with a stable period of constant rate $\delta_2 < -3/2$ for the PBH formation imposes an additional constraint across the scaling dimensions. For PBH scenarios that realize a continuous decay of the power spectrum to the end of inflation (mostly for the generation of PBHs in a certain mass range), $\Delta_3 = -\delta_3$ is constrained by the dimension $\Delta_2 = 3 +\delta_2$ (for $\delta_2 < -3$) of the negative rate so that $\delta_3 \rightarrow -3-\delta_2$. On the other hand, if one intends to realize a secondary slow-roll phase with $\Delta_3 \approx 0$, the continuity $\Delta_3 = \Delta_2$ implies that the negative rate always goes to the ultra-slow-roll limit where $\delta_2 \rightarrow -3$.  
\footnote{We thank also the numerical confirmations of this property from H. V. Ragavendra and J. Silk in their study \cite{Ragavendra:2020sop}.}

\subsection{$N$-stage inflation}\label{Sec. N_stage}
The unified mode function derived in Section~\ref{Sec. bulk3} can be extended to more general cases with $N > 3$ phases of inflation (see also \cite{Byrnes:2018txb}), as long as the rolling rates, $\delta_N$, in the $N$-th phase of inflation are nearly constant and the background spacetime is sufficiently close to the de Sitter phase ($\epsilon_N \ll 1$) in each phase. One of the possible application for the $N$-stage constant-rate inflation is to generate multiple peaks in the power spectrum for producing PBHs in different mass ranges \cite{Tada:2019amh}. 

Although the $N$-stage extension may largely increase the complexity of the final mode functions at the end of inflation, there are some generic features shared among the boundary arguments that shall be outlined in this section:
\begin{enumerate}
	\item \textit{The final spectrum.} The instantaneous transition of rolling rates to $\delta_N$ is equivalent to altering only the effective mass for the Mukhanov-Sasaki variable $v_N$ in the $N$-th phase. Therefore the solution of $v_N$ takes the similar structure as
	\begin{align}\label{MS_variable_N}
	v_N = c_N^{(1)} \sqrt{\tau_N} H_{\nu_N}^{(1)} (\tau_N) +  c_N^{(2)} \sqrt{\tau_N} H_{\nu_N}^{(2)} (\tau_N),
	\end{align}
	where $\nu_N = \vert 3/2 + \delta_N\vert$ and $\tau_N = -k \eta$. At the end of inflation, $\tau_{\rm end} = k/k_{\rm end} = \tau_\ast x_2x_3\cdots x_N$, where $x_N = k_{\ast(N-1)}/k_{\rm end} \approx a_{\ast(N-1)}/a_{\rm end} = e^{-\Delta N_{\ast(N-1)}}$ measures the ratio of the horizon scale between the end of $N-1$-th stage to the end of $N$-th stage. The final power spectrum for $\tau_{\rm end} \ll 1$ modes is then
	\begin{align}\label{P_R_N_stage}
	P_{\mathcal{R}}^{\rm IR}(k; \eta_{\rm end}) = \frac{H_k^2}{8\pi} \frac{1}{\epsilon_{\rm end}} 
	\left\vert f_{\mathcal{R}}^N (\tau_\ast, x_2, \cdots, x_N)\right\vert^2
	\left\vert \frac{i}{\pi} 2^{\nu_N} \Gamma(\nu_N)\right\vert^2
	\left(\frac{k}{k_{\rm end}}\right)^{2\Delta_N},
	\end{align}
	where $\Delta_N = 3/2 -\nu_N$, $f_{\mathcal{R}}^N \equiv  (c_N^{(2)} - c_N^{(1)} )/c_1$ and $\epsilon_{\rm end} = \epsilon_\ast x_2^{-2\delta_2}x_3^{-2\delta_3}\cdots x_N^{-2\delta_N}$.

	\item\textit{The large $k$ limit.} The mode function \eqref{MS_variable_N} has the general property $\sqrt{\tau_N}H_{\nu_N}^{(1,2)}(\tau_N) \rightarrow  \sqrt{2/\pi} e^{\mp i \frac{\pi}{2}(\nu_N+\frac{1}{2})} e^{\pm i \tau_N}$ in the limit of $k \rightarrow \infty$ where $\tau_N \rightarrow \infty$. This leads to the matching of the boundary conditions at each phase transition, $\eta_{\ast N} = -1/k_{\ast N}$ with $k/k_{\ast N} \gg \frac{\pi}{2}(\nu_N+\frac{1}{2})$, which is approximately given by the relation
	\begin{align}
	\left(c_N^{(2)}+c_N^{(1)}\right) &\cos \left( -k/k_{\ast N}\right)+ \left(c_N^{(2)} -c_N^{(1)}\right) i \sin \left( -k/k_{\ast N}\right) 
	\\\nonumber
	&=  \left(c_{N+1}^{(2)}+c_{N+1}^{(1)}\right) \cos \left( -k/k_{\ast N+1}\right)+ \left(c_{N+1}^{(2)} -c_{N+1}^{(1)}\right) i \sin \left( -k/k_{\ast N+1}\right), 
	\end{align}
	which implies that $c_{N+1}^{(2)} -c_{N+1}^{(1)} \rightarrow c_N^{(2)} -c_N^{(1)}$.
	 The coefficient $c_1$ of the primary slow-roll phase fixed by the Bunch-Davies vacuum \eqref{c_1} ensures that $  f_{\mathcal{R}}^2 = (c_2^{(2)} - c_2^{(1)} )/c_1$ is a pure oscillatory function of $k$ (no power-law dependence) in the limit of $k \rightarrow \infty$. One can repeat the matching process to find that $ f_{\mathcal{R}}^N \equiv (c_N^{(2)} -c_N^{(1)})/c_1$ is also a pure oscillatory function of $k$, and thus the final spectrum \eqref{P_R_N_stage} has the asymptotic scaling dimension $	P_{\mathcal{R}}^{\rm IR}(k\gg k_{\ast (N-1)}; \eta_{\rm end}) \sim k^{2\Delta_N}$. 
	
	\item\textit{Continuous scaling.} For the $N$-th scaling dimension $\vert \Delta_{N} \vert\gg 0$, the boundary argument in Section~\ref{Sec. boundary12} and \ref{Sec. boundary23} indicate that $k^{3/2}\mathcal{R}_{N+1} \sim \mathcal{C}_{N0} k^{\Delta_{N}}$ (even with $\vert\Delta_{N+1}\vert \rightarrow 0$) and $k^{3/2}\mathcal{R}_{N+1} \rightarrow k^{\Delta_{N}}$ when $\tau_\ast \rightarrow 0$. This means that there is a short period of continuous scaling $P_{\mathcal{R}} \sim k^{2\Delta_{N}}$ for $k \gtrsim k_{\ast N}$ and the continuous scaling breaks down to $P_{\mathcal{R}} \sim k^{2\Delta_{N+1}}$ when $k \gg k_{\ast N}$ due to the large $k$ behavior discussed above. 
	
	\item\textit{Steepest growth.} For PBH scenarios that allow multiple peaks in the power spectrum, the $P_{\mathcal{R}} \sim k^4$ growth appears whenever a slow-roll phase ($\Delta_{N -1} \rightarrow 0$) transits to a negative-constant-rate phase ($\Delta_{N} = 3+\delta_N$ with $\delta_N < -3/2$), as argued in Section~\ref{Sec. boundary12}. To reduce the power spectrum from the peak values after enhancement, we need $\Delta_{N+1} = -\delta_{N+1}$ to be a positive-constant-rate phase with $\delta_{N+1} >0$. To maintain a stable phase with $\delta_N < -3$, the inflaton potential should satisfy the condition $\delta_{N+1} = -3-\delta_N$ as numerically shown in Section~\ref{Sec. PBH scenario}.
\end{enumerate}

\section{Criterion of conformal weight violation}\label{Sec:criterion}
In the slow-roll ($\delta = 0$) and the ultra-slow-roll ($\delta = -3$) phases, the scaling dimension, $\Delta = 3/2- \vert 3/2 + \delta\vert =0$, vanishes identically. A question arises following this fact is that, why the $P_{\mathcal{R}} \sim k^4$ growth of the power spectrum only occurs when inflation transits from the slow-roll (SR) phase to the ultra-slow-roll (USR) phase, yet the $k^4$ growth does not occur in the inverse (USR $\rightarrow$ SR) transition? (In fact, $P_\mathcal{R}$ keeps scale-invariant for the USR to SR transition, see \cite{Cai:2017bxr}.) In this section, we show that the $k^4$ growth as a conformal-weight-violating process is due to the entropy production of the inflaton perturbation driven by the sharp deceleration of the coherent inflaton motion in Phase 2.

The criterion for realizing an enhanced power spectrum through a decay of the first slow-roll parameter $\epsilon_H \simeq \dot{\phi}^2/(2M_{\rm pl}^2H_\ast^2)$ has been given in \cite{Leach:2001zf}, and the connection with entropy production for such an enhancement in USR inflation has been pointed out in \cite{Ragavendra:2020sop,Leach:2000yw}. Here, we apply the discussion to the generic constant-rate inflation to identify the origin of the $k^4$ growth. Our formula for the inflaton perturbation is obtained in the spatially flat slicing of the metric:
\begin{align}\label{metric:flat_slicing}
ds^2 =  (1+\alpha)^2 dt^2 + a^2h_{ij} \left(dx^i + \beta^i dt \right) \left(dx^j + \beta^j dt \right),
\end{align}   
where $\alpha \equiv N- 1$, $\beta^i \equiv N^i$, and $N$ ($N^i$) is nothing but the lapse (shift) function in the standard Arnowitt-Deser-Minser (ADM) formalism. We simply take $h_{ij} = \delta_{ij}$ since the tensor perturbation is irrelevant to our discussion.
At the linear order, the Einstein equation for the energy density, pressure, and momentum flux gives the definition to the inflaton fluid perturbation as
\begin{align}\label{def:delta_phi}
\delta\rho_\phi &= \dot{\phi}\left(\delta\dot{\phi} -\alpha \dot{\phi}\right) + V_\phi \delta\phi, \\
\delta p_\phi &= \dot{\phi}\left(\delta\dot{\phi} -\alpha \dot{\phi}\right) - V_\phi \delta\phi, \\
\delta q_\phi &= -\dot{\phi} \delta\phi,
\end{align}
where $\delta\rho_m \equiv \delta\rho_\phi - 3H\delta q_\phi$ represents the gauge-invariant total matter perturbation \cite{Gordon:2000hv}. Together with the vanishing of the anisotropic stress for the linear inflaton perturbation, one can solve $\alpha$ and $\beta$ in terms of $\delta\phi$ as
\begin{align}\label{sol:lapse_shift}
\alpha = \frac{\dot{\phi}}{2H M_{\rm pl}^2} \delta\phi, \qquad
\partial^2\beta = \frac{\dot{\phi}^2}{2H^2 M_{\rm pl}^2} \frac{d}{dt} \left(-\frac{H}{\dot{\phi}} \delta\phi\right),
\end{align}
where $\beta$ is the scalar mode of $\beta^i$ with respect to the decomposition $\beta^i = \partial_i \beta + \beta_T^i$ and $\partial_i\beta_T^i = 0$.
These are all the quantities we need for the discussion of entropy production.

\bigbreak
\noindent
\textbf{Inflaton mode functions.}
The constant-rate condition illustrated in Figure~\ref{fig:phase_plot} in fact specifies the first and second field derivatives of the inflaton potential $V(\phi)$ so that one can solve the mode functions of $\delta\phi$ based on the given parameter $\delta$. 
Let us derive the analytic solutions of $\delta\phi_k$ before going to the fluid dynamics of the inflaton field.

Recalling that the coherent motion of $\phi$ follows $\ddot{\phi} + 3H\dot{\phi} + V_\phi =0$, which can be translated to $\delta + 3 = -V_\phi/(H\dot{\phi})$ in terms of the rate-of-rolling parameter $\delta$. Taking the time derivative of $\delta$, one finds
\begin{align}\label{CR condition}
\frac{\dot{\delta}}{H} = -\frac{V_{\phi\phi}}{H^2} + (\epsilon- \delta) (\delta + 3), 
\end{align} 
where the constant-rate condition $\dot{\delta} =0$ with $\epsilon \ll \delta$ indicates a constant inflaton mass $V_{\phi\phi}  = -\delta(\delta + 3)H^2$. 

On the other hand, the equation of motion of $\delta\phi$ in the flat slicing \eqref{metric:flat_slicing} is given by
\begin{align}
\delta\phi^{\prime\prime} + 2\mathcal{H}\delta\phi^\prime +\left(k^2 + a^2 V_{\phi\phi}\right) \delta\phi 
+ 2\left( \phi^{\prime\prime} + 2\mathcal{H}\phi^\prime\right)\alpha + \phi^\prime \left(\alpha^\prime +a \partial^2\beta\right) = 0,
\end{align}
where the prime denotes the derivative with respect to the conformal time.
One can check from the solutions \eqref{sol:lapse_shift} to find that those terms involved with $\alpha$ or $\beta$ are suppressed by $\epsilon_H$ in the de Sitter background. Therefore, using the constant-rate condition \eqref{CR condition} and dropping $\epsilon_H$ suppressed terms, the leading equation of motion for the mode function $u_k = a \delta\phi_k$ reads
\begin{align}
\frac{\partial^2 u}{\partial z^2} + \left[1 - \frac{2+ \delta (\delta +3)}{z^2} \right] u = 0,
\end{align}
which is nothing but the equation of motion for the mode function $v_k$ of the Mukhanov-Sasaki variable given in \eqref{eom:MS_1} and \eqref{eom:MS_2} with $\nu \equiv \vert \delta + 3/2\vert$. This is not surprising since we are reproducing the standard expression for the curvature perturbation in flat slicing \cite{Maldacena:2002vr,Gordon:2000hv}:
\begin{align}\label{def:R_flat_slicing}
\mathcal{R} = - \frac{H}{\dot{\phi}} \delta\phi.
\end{align}
The solution \eqref{sol:lapse_shift} is therefore $\partial^2\beta = \epsilon_H \dot{\mathcal{R}}$.

\bigbreak
\noindent
\textbf{Entropy production.}
The entropy perturbation in single-field inflation is usually negligible. However, the excitation of both constants of integration, $c_N^{(1,2)}$, due to the transition into a negative rolling rate in Phase 2, indicates that the late-time curvature perturbation contains both adiabatic and entropy modes. It is the exponential growth of the entropy mode with $\delta_2 < -3/2$ that sources the enhancement of $P_\mathcal{R}$ on superhorizon scales \cite{Leach:2000yw}. A numerical experiment to demonstrate the growth of superhorizon modes can be found in \cite{Cheng:2018qof}. To see the role of entropy perturbation in the enhancement of $P_\mathcal{R}$, let us recall the gauge-invariant definition for a single scalar fluid \cite{Gordon:2000hv} as 
\begin{align}
S_\phi = H\left(\frac{\delta p_\phi}{\dot{p}_\phi} - \frac{\delta \rho_\phi}{\dot{\rho}_\phi}\right),
\end{align}
where in constant-rate inflation $\dot{\rho}_\phi = -3H\dot{\phi}^2$ and $\dot{p}_\phi = (2\delta + 3)H\dot{\phi}^2$. Taking the perturbations given by \eqref{def:delta_phi}, one can manipulate the definition to find that
\begin{align}
S_\phi = \frac{2(\delta + 3)}{3(2\delta + 3)} \frac{1}{\dot{\phi}}  \left( \delta\dot{\phi} - \alpha\dot{\phi} -\delta H \delta\phi \right) 
= \frac{2(\delta + 3)}{3(2\delta + 3)} \frac{\delta\rho_m}{\dot{\phi}^2} .
\end{align}
Given that the total matter perturbation $\delta\rho_m = 2 H M_{\rm pl}^2 \partial^2\beta$ in flat slicing, we can connect the time-derivative of the curvature perturbation \eqref{def:R_flat_slicing} to the entropy perturbation as
\begin{align}
\dot{\mathcal{R}} = -\frac{H}{\dot{\phi}}  \left( \delta\dot{\phi} - \alpha\dot{\phi} -\delta H \delta\phi \right)  \equiv -\frac{3(2\delta + 3)}{2(\delta + 3)} H S_\phi.
\end{align}
This relation shows the generic suppression of superhorizon evolution of the curvature perturbation in slow-roll inflation ($\delta \rightarrow 0$) as $d\mathcal{R}/dN \sim S_\phi \sim (k/aH)^2 \ll 1$. 

The above definition of the entropy perturbation can be misleading to a constant $\mathcal{R}$ in the exact USR inflation with $\delta \rightarrow -3$ where $\dot{\mathcal{R}} = S_\phi \rightarrow 0$. However, non-zero entropy has been created in the primary slow-roll phase (Phase 1) and its change in time is governed by 
\begin{align}
\dot{S}_\phi = -\left(3 + \epsilon_H + 2\delta \right) H S_\phi + \frac{2(\delta + 3)}{3(2\delta + 3)} \frac{k^2}{a^2}\frac{\mathcal{R}}{H},
\end{align}
where we have applied the constant-rate condition \eqref{CR condition} to simplify the expression. For a consecutive phase with $\delta < -3/2$, one finds $\dot{S}_\phi \approx -(3+2\delta)H S_\phi$ leading to a rapid growth $S_\phi \sim e^{-(3+2\delta)N}$ on large scales with $k \ll a H$, which allows the subdominant entropy mode to overcome the adiabatic perturbation.

The relation $\dot{\mathcal{R}} =\partial^2\beta/\epsilon_H \sim H S_\phi$ provides an alternative way to see that the late-time curvature perturbation contributed by the entropy production has the scaling property $\mathcal{R}_S \sim k^2$, where $\mathcal{R}_S$ comes from the integration 
\begin{align}\label{def:R_AS_decomposition}
\mathcal{R} \equiv \mathcal{R}_A + \mathcal{R}_S = \mathcal{R}_A + k^2 \int \frac{\beta(t)}{\epsilon_H(t)} dt.
\end{align}
$\mathcal{R}_A$ is nothing but the adiabatic mode of the curvature perturbation, which is a constant in the slow-roll and the USR phases. $\mathcal{R}_S \geq \mathcal{R}_A$ is realized when $\epsilon_H$ decays to a very small value with $\delta < -3/2$, which summarizes the criterion of superhorizon enhancement given in \cite{Leach:2001zf}. Comparing the adiabatic-entropy decomposition \eqref{def:R_AS_decomposition} with the boundary argument in Section~\ref{Sec. boundary12}, we find that $\mathcal{R}_S \sim k^{\Delta +2}$ is the subleading mode with the next-to-lowest conformal weight.
As a result, the temporary domination of $\mathcal{R}_S$ leads to a growing spectrum $P_\mathcal{R} \sim k^4$ and breaks the continuity of the leading conformal weight of the system.
\footnote{The $k^4$ growth reflects the superhorizon evolution of the curvature perturbation for modes (in the range of $k_{\rm min} < k < k_\ast$) that have exited the horizon close to the end of Phase 1. This superhorizon evolution is induced by the deceleration dynamics of inflaton (namely $\delta_2 < 0$) in Phase 2. The larger the spectral amplitude is enhanced at $k = k_\ast$, the smaller the wavenumber $k_{\rm min}$ needs to be (see Figure~\ref{fig:spectrum_phase2}). For modes that exit the horizon well inside Phase 2 or in Phase 3 with $\delta_3 = -3-\delta_2$ (that is for $k_\ast \ll k < k_{\rm end}$), the spectral amplitude at the end of inflation is governed by two factors related to the Phase 2 parameters as $P_\mathcal{R}(k > k_\ast; \eta_{\rm end}) \sim (k_{\ast}/k_{\ast\ast})^{2\delta_2}(k/k_{\ast\ast})^{6+2\delta_2}$, as shown in \eqref{P_R_phase3_IR} or \eqref{P_R_IR}. The scale-independent factor $ (k_{\ast}/k_{\ast\ast})^{2\delta_2}$ is solely determined by the duration of Phase 2, namely $\Delta N = \ln(k_{\ast\ast}/k_\ast) = \ln(\eta_{\ast}/\eta_{\ast\ast})$, due to the decay of the first slow-roll parameter $\epsilon_2(\eta_{\ast\ast}) = \epsilon_\ast (\eta_{\ast}/\eta_{\ast\ast})^{-2\delta_2} = \epsilon_\ast e^{2\delta_2 \Delta N}$. The scale-dependent factor $(k/k_{\ast\ast})^{6+2\delta_2}$ describes the momentum scaling on the final boundary (fixed by the conformal weight: $6+ 2\delta_2 = 2\Delta_2 = 2\Delta_3$) converted from the superhorizon (time) evolution of the curvature perturbation with respect to the dilatation symmetry. For the transient USR case with $\delta_2 = -3$, the decay of $\epsilon$ during Phase 2, factorized by $(k_{\ast}/k_{\ast\ast})^{2\delta_2}$, is the only enhancement of the power spectrum $P_\mathcal{R}$.}

On the other hand, in a Phase $N$ with a decaying $P_\mathcal{R}$ driven by a positive rate $\delta_N > 0$, the parameter $\epsilon_H$ is growing with time so that the entropy mode $\mathcal{R}_S$ never dominates the adiabatic perturbation. This is the reason why the conformal weight of $\mathcal{R}$ keeps unchanged from the previous phase ($\Delta_{N} = \Delta_{N -1}$), where the constraint $\delta_{N} = -3 -\delta_{N - 1}$ for $\delta_{N -1} \leq -3$ is protected by the adiabatic condition.  

To answer the question at the beginning of this section, let us denote the conjugate momentum of the curvature perturbation $\mathcal{R}$ in a dimensionless way based on the general mode function \eqref{MS_variable_N} as 
\begin{align}
\pi_N^{\mathcal{R}} \equiv \frac{\partial \mathcal{R}_N}{\partial z} 
= - A_N \frac{\mathcal{R}_N}{z}  
+\frac{\sqrt{-z}}{y_N} \left[c_N^{(1)}H_{\nu_N-1}^{(1)}(-z) + c_N^{(2)}H_{\nu_N - 1}^{(2)}(-z)\right],
\end{align}
where $A_N\equiv \Delta_{N}+\delta_N = \epsilon_{N,2} + \Delta_{N}$ is nothing but the $A_i$ appears in \eqref{boundary_constraint_2} with $i = N$ and $\epsilon_{i} \rightarrow 0$. Given that $\Delta_{N} = 3/2 -\vert 3/2 + \delta_N \vert$, one finds two possibilities as
\begin{align}
A_N = \Delta_{N}+\delta_N = 
\left\{ 
\begin{array}{lll}
\;0, &\quad \delta_N \geq -3/2,  & \quad (\textrm{adiabatic})\\
\;2\delta_N + 3, &\quad \delta_N < -3/2 & \quad (\textrm{non-adiabatic}).
\end{array}
\right.
\end{align}
We shall refer the case with $A_N \neq 0$ as the ``non-adiabatic'' phase for the curvature perturbation with $k^{3/2}\mathcal{R}_N \sim c_N^{(1,2)} (-z)^{\Delta_{N}}$ in the late-time limit, where the leading scaling power of the conjugate momentum is simply $k^{3/2}\pi_N^{\mathcal{R}} \sim k^{3/2}A_N \mathcal{R}_N /z \sim  c_N^{(1,2)} (-z)^{\Delta_{N}-1}$. On the other hand, the case with
$A_N = 0$ is referred as the ``adiabatic'' phase where $k^{3/2}\pi_N^{\mathcal{R}} \sim c_N^{(1,2)} (-z)^{1+\Delta_{N}}$ and the conjugate momentum for the leading mode in $\mathcal{R}_N$ is missing.

It is now clear that the transition from an adiabatic Phase $N$ to a non-adiabatic Phase $N+1$ has to respect the continuity of the conjugate momentum at $z = z_{\ast N}$, where $c_N^{(1,2)} (-z_{\ast N})^{1+\Delta_{N}} \sim c_{N+1}^{(1,2)} (-z_{\ast N})^{\Delta_{N +1} - 1}$ implies $c_{N+1}^{(1,2)} \sim (-z_{\ast N})^{2 + \Delta_{N} - \Delta_{N +1}}$ in the large $k$ limit for Phase $N$, see Section~\ref{Sec. N_stage}. This gives 
\begin{align}
k^{3/2}\mathcal{R}_{N+1} (-z) \sim c_{N+1}^{(1,2)} (-z)^{\Delta_{N +1}} \sim (k/k_{\ast N})^{2+\Delta_{N}}
\end{align}
when seeing from a late-time boundary $-z \ll -z_{\ast N}$. As a result, a $P_\mathcal{R} \sim k^4$ growth appears with $\Delta_{N} \rightarrow 0$ due to the entropy production across the transition. This is indeed the case for the SR to USR transition where $\Delta_{N} = \Delta_{N +1} = 0$ but $A_N =0$, $A_{N+ 1} = -3$.

For a non-adiabatic to adiabatic transition (such as the USR to SR transition \cite{Cai:2017bxr}), one can repeat the above process to find that the continuity of the conjugate momentum is led by $k^{3/2}\mathcal{R}_{N+1} (-z) \sim (k/k_{\ast N})^{\Delta_{N} -2}$, which in fact describes a decaying mode since the range holds only for $k/k_{\ast N} \lesssim 1$. In this case the leading term comes from the matching of $\mathcal{R}$ at the phase transition and the conformal weight is inherited from the adiabatic mode $\mathcal{R}_A$ in the preceding phase, as argued in Section~\ref{Sec. boundary23} and \ref{Sec. bulk3}. Finally, the continuity of $\pi_{N+1}^{\mathcal{R}}$ from an adiabatic to adiabatic (or a non-adiabatic to non-adiabatic) transition gives $c_{N+1}^{(1,2)} \sim (k/k_{\ast N})^{\Delta_{N} - \Delta_{N +1}}$, which is the same result as that obtained from the continuity of $\mathcal{R}$ matching.

In summary, the conformal weight violation driven by a temporary entropy domination only occurs for an adiabatic to non-adiabatic transition.

\section{Conclusions}\label{Sec. conclusion}

In this work, we have shown that the scaling dimensions (or conformal weights) fixed by the dilatation symmetry of the de Sitter background can be a powerful tool to understand the final power spectrum at the end of inflation for a majority classes of PBH scenarios that are involved with enhanced spectral amplitudes led by a phase with the rolling rate of the inflaton field being negative and nearly constant. 
The continuity of the conformal weights across different scaling dimensionalities imposes a constraint to the inflaton rolling rate in the PBH scenarios, where the spectral shape on scales towards the end of inflation can be a power-law decay or scale-invariant. 
The apparent violation of the continuity across conformal weights from the (primary) slow-roll phase to the negative-constant-rate phase can be realized by the efficient entropy production in the PBH scenarios.
Our analytic formalism has identified the entropy perturbation as the subleading mode carrying the next-to-lowest conformal weights and its temporary domination triggers the $k^4$ growth of the power spectrum, which is the answer to the steepest growth problem \cite{Byrnes:2018txb}. 
We have provided a generalized adiabatic condition in terms of the conformal weight $\Delta$ and the rolling rate $\delta$ for the constant-rate inflation.
Despite that those PBH scenarios in general require fine tuning the model parameters, it is interesting to consider them as the realization of conformal field systems that experience the time-varying conformal weights during inflation. 

Further study is required to clarify the necessary conditions between adiabaticity and the continuity across scaling dimensions. Revisiting the problem via the quantization of scalar fields in terms of the hyperbolic coordinates for the $3 +1$ de Sitter space \cite{Strominger:2001pn,Ng:2012xp,Jafferis:2013qia,Brown:2017osf} is one of the conceivable working directions. It is also interesting to seek for possible signatures, due to the time-varying conformal weights, residing in higher-order or non-Gaussian correlation functions of the curvature perturbation. The presented analytic formulae could be applied as a zeroth-order result to study subhorizon corrections to the superhorizon correlators arisen from the off-attractor nature of the constant-rate inflation \cite{Biagetti:2018pjj,Vennin:2020kng,Ando:2020fjm,Cai:2017bxr,Suyama:2021adn,Pattison:2021gpv,Pattison:2021oen}.

\acknowledgments
The authors thank Xingang Chen, Kalliopi Petraki, H. V. Ragavendra, and Joseph Silk for their helpful comments and discussions.
K.-W. Ng is supported in part by the Ministry of Science and Technology (MOST) of Taiwan, R.O.C. under Grant No. MOST 109-2112-M-001-003.
Y.-P. Wu is supported by the the Agence Nationale de la Recherche (ANR) Accueil de Chercheurs de Haut
Niveau (ACHN) 2015 grant (“TheIntricateDark” project).



\end{document}